\documentclass[twocolumn,letter]{aastex701}

\usepackage{graphicx}
\usepackage{graphbox}
\usepackage{hyperref}
\usepackage{amssymb,amsmath}
\usepackage{float}

\newcommand{\gtorder}{\mathrel{\raise.3ex\hbox{$>$}\mkern-14mu\lower0.6ex\hbox{$\sim$}}}
\newcommand{\ltorder}{\mathrel{\raise.3ex\hbox{$<$}\mkern-14mu\lower0.6ex\hbox{$\sim$}}}

\newcommand{\nk}[1]{{\color{red} \bf #1}}

\begin{document}

\title{A Possible X-ray and Gamma-ray Quasi-Periodic Oscillation in GRB 241030A}

\author[0000-0002-7465-0941]{Noel Klingler}
\affiliation{Center for Space Sciences and Technology, University of Maryland, Baltimore County, Baltimore, MD, 21250, USA}
\affiliation{Astroparticle Physics Laboratory, Astrophysics Science Division, NASA Goddard Space Flight Center, Greenbelt, MD 20771, USA}
\affiliation{Center for Research and Exploration in Space Science and Technology, NASA Goddard Space Flight Center, Greenbelt, MD 20771, USA}
\email{noelklin@umbc.edu}

\author[0000-0003-2759-1368]{Cecilia Chirenti}
\affiliation{Department of Astronomy, University of Maryland, College Park, MD 20742-2421, USA}
\affiliation{Astroparticle Physics Laboratory, Astrophysics Science Division, NASA Goddard Space Flight Center, Greenbelt, MD 20771, USA}
\affiliation{Center for Research and Exploration in Space Science and Technology, NASA Goddard Space Flight Center, Greenbelt, MD 20771, USA}
\email{chirenti@umd.edu}

\author[0000-0002-2666-728X]{M.~Coleman~Miller}
\affiliation{Department of Astronomy and Joint Space-Science Institute, University of Maryland, College Park, MD 20742-2421 USA}
\email{miller@astro.umd.edu}

\author[0000-0002-7851-9756]{Amy Lien}
\affiliation{University of Tampa, Department of Physics and Astronomy, 401 W. Kennedy Blvd, Tampa, FL 33606, USA}
\email{alien@ut.edu}

\author[0000-0001-6849-1270]{Simone Dichiara}
\affiliation{Department of Astronomy and Astrophysics, The Pennsylvania State University, 525 Davey Lab, University Park, PA 16802, USA}
\email{sbd5667@psu.edu}


\begin{abstract}

Quasiperiodic oscillations (QPOs) in gamma-ray intensity have been reported from a few gamma-ray bursts.  
These QPOs could be related to fundamental frequencies in the sources, which would lead to new insights about these systems.  Here we report on our analysis of GRB~241030A, a several-minute-long burst with $\sim 10$ gamma-ray flares that have approximate periodicities at $f \sim 0.04$~Hz and $f\sim 0.08$~Hz.  
This quasi-periodicity is seen in X-rays with the Swift X-Ray Telescope (XRT) as well as in gamma-rays with the Swift Burst Alert Telescope (BAT) and the Fermi Gamma-ray Burst Monitor (GBM). 
Compared with a power spectral model that has only red noise, the Bayes factor in favor of a Lorentzian-shaped QPO is more than $10^{200}$, even when the red noise is described by segments with up to four different slopes.  
However, the interpretation of this power spectral feature as a QPO rather than as a fluctuation in red noise is complicated by the low frequency of the feature as well as the extremely large number of counts, which amplifies differences from smooth power spectral distributions.  
After discussing the properties of the apparent QPO, we discuss a few processes that could plausibly modulate the flux at this frequency. 
Among the mechanisms considered, we find the most plausible explanation to be a GRB jet passing through regularly spaced shells of circumburst material, possibly produced by binary interactions near periastron in an eccentric progenitor system.

\end{abstract}


\section{Introduction} \label{sec:Intro}

The extreme energies and rapid fluctuation times of gamma-ray bursts (GRBs) mean that their study contacts numerous aspects of fundamental physics and astrophysics, including plasma physics, nuclear physics, and general relativity.  
A consensus has emerged that GRBs fall into two broad phenomenological classes \citep{1993ApJ...413L.101K}: ``short'' bursts ($T_{90}\lesssim 2$~s, where $T_{90}$ is defined as the interval between the time when 5\% of the counts above background have been accumulated, and the time when 95\% of the counts above background have been accumulated), generally thought to arise from compact-object mergers involving two neutron stars or a neutron star and a black hole \citep{1989Natur.340..126E,2017ApJ...848L..12A}, and ``long'' bursts ($T_{90}\gtrsim 2$~s), which are associated with at least some core-collapse supernovae \citep{1993ApJ...405..273W,2003Natur.423..847H,2003ApJ...591L..17S}.
Category-blurring GRBs do exist, e.g., the ``long short burst" GRB~211211A \citep{2022Natur.612..223R}, but these two categories are relatively robust.

In recent years, increasing attention has been devoted to quasi-periodic oscillations (QPOs) in GRBs, in which the flux or count rate exhibits variability at one or more characteristic frequencies. 
The lure of this phenomenon is that if a QPO can be established with confidence, then it could correspond to a characteristic frequency of the central engine, which would then serve as a probe of the physics of ultra-dense matter, relativistic outflows, and/or hyper-accreting objects.  
For example, \citet{2023Natur.613..253C} associated the $\sim 2600$~Hz QPO that they found in two Burst and Transient Source Experiment (BATSE) bursts with the $l=2$ mode seen in numerical simulations of mergers, and the $\sim 1000$~Hz QPO in those bursts with a quasispherical oscillation of the remnant.  
Given these associations, \citet{2025ApJ...983...88G} constrained the circumferential radius of a canonical $M=1.4~M_\odot$ neutron star to be $\approx 12.5\pm 0.4$~km. Although this is model-dependent, it is as tight as any radius constraint and is consistent with the radius range inferred using other methods.

QPOs have been suggested in a number of other bursts as well (including but not limited to \citet{2009Ap.....52..534D,2014MNRAS.441.2375H,2022MNRAS.513L..89Z,2024ApJ...967...26C,2024ApJ...973..126Z,2025ApJ...985...33G,2025A&A...702A.149H,2025ApJ...994...21S,2025MNRAS.537.2313Y,2025MNRAS.541.3787S,2026ApJ...998..289C}).  
It is often challenging to assess the probability that these QPOs are real for two major reasons: (a)~searches often involve hidden trials factors (i.e., there are many choices of possible frequencies, start and end times of the segment analyzed, and energy channels), and (b)~because GRBs have sharp flux variations it is often difficult to decouple ``normal'' variation from QPO variation.  
Moreover, especially if the suggested frequency is relatively low, even if a QPO has been established statistically its interpretation could remain murky. 

Here we present an analysis of the first few hundred seconds of data from the long burst GRB~241030A. 
Swift/BAT, Swift/XRT, and Fermi/GBM data all show a $\sim 0.04$~Hz and a $\sim0.08$~Hz feature in the power spectrum.  
This is a promising signal, but as we discuss below, this and similar bursts often show substantial low-frequency complexity and thus caution is warranted. 
In Section~\ref{sec:data} we describe the GRB's observed characteristics. 
In Section~\ref{sec:gamma} we discuss data analysis and our search for QPOs in the data. 
In Section~\ref{sec:discussion} we discuss possible physical interpretations if the QPO represents a characteristic frequency of the system, and in Section \ref{sec:conclusions} we present our conclusions.

\section{GRB 241030A}
\label{sec:data}

At 05:48:03 UT on 30 October 2024, GRB 241030A was detected simultaneously with the Fermi Gamma-ray Burst Monitor (GBM; trigger 751960088.32731 / 241030242; \citealt{2024GCN.37955....1F}), the Fermi Large Area Telescope \citep[LAT;][]{2024GCN.37979....12}, and the Swift Burst Alert Telescope (BAT; trigger 1263718; \citealt{2024GCN.37956....1K}). 
We use the GBM $T_0$ (05:48:03.33 UT) as our reference point; the BAT $T_0$ was virtually the same (05:48:03.12 UT). 

\subsection{Fermi/GBM}

Fermi/GBM consists of 12 sodium iodide (NaI) and two bismuth germanate (BGO) scintillation detectors. 
The arrangement of these detectors allows GBM to observe the entire unocculted sky (field of view [FoV] $\sim 8$ sr). 
The NaI and BGO detectors cover energy ranges of $\sim 8 \, {\rm keV} - 1 \, {\rm MeV}$ and $\sim 200 \, {\rm keV} - 40 \, {\rm MeV}$, respectively \citep{2009ApJ...702..791M}.

GRB 241030A  mostly illuminated the n0 and n1 NaI detectors and the b0 BGO detector of Fermi/GBM.
The burst was also identified with the Fermi/LAT, at $\sim$16$^{\circ}$ from the instrument boresight with a photon flux above 100 MeV of (4.0 $\pm$ 1.0) $\times$ 10$^{-6}$ cts cm$^{-2}$ s$^{-1}$ in the time interval from $T_0$ to $T_0+800$~s \citep{2024GCN.37979....12}.
The GBM light curve consists of two main emission episodes, one from $\sim T_0$ to $\sim T_0+50$ s and another one from $\sim T_0+100$ s to $\sim T_0+200$ s (see Figure \ref{fig:tdrss_BAT_LC}, top panel). 
Each of the episodes consists of multiple pulses. 
For this burst $T_{90}$ is $\sim 166$~s in the $50-300$~keV energy range.

The best-fit spectrum reported in the GBM online catalog\footnote{https://heasarc.gsfc.nasa.gov/W3Browse/fermi/fermigbrst.html}, extracted over the interval from $T_{0}-1.024$~s to $T_{0}+225.280$~s, is well described by a power law with an exponential cutoff \citep[``Comptonized'' model;][]{2014ApJS..211...13V, 2014ApJS..211...12G, 2016ApJS..223...28N, 2020ApJ...893...46V}. The fitted spectral photon index is $-1.279 \pm 0.007$ and the peak energy is $E_{\rm peak} = 152.4 \pm 3.5$ keV. This model yields a total fluence of $(5.12 \pm 0.08) \times 10^{-5}$~erg~cm$^{-2}$ over the same time interval, in the energy range $10-1000$~keV.

\subsection{Swift/BAT}
\begin{figure}[ht]
\centering
\includegraphics[width=0.46\textwidth]{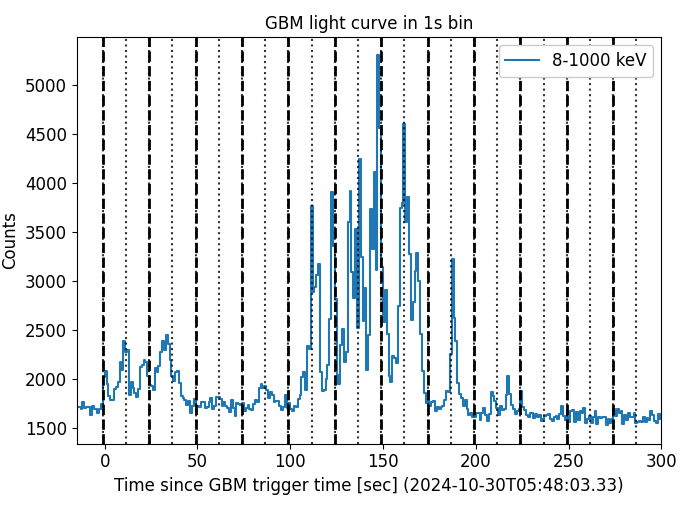}
\includegraphics[width=0.5\textwidth]{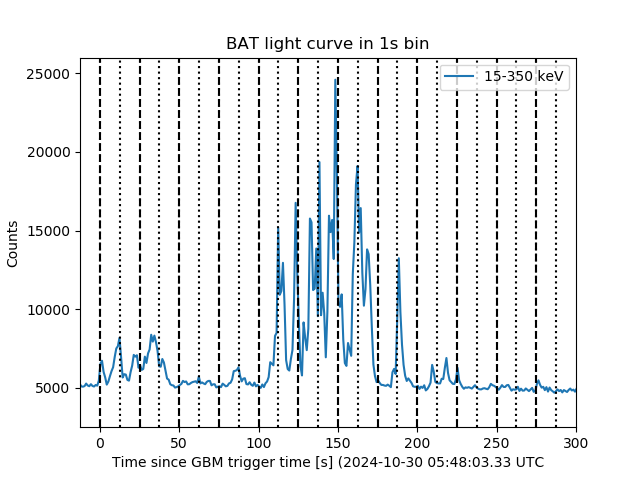}
\caption{
{\sl Top:} Fermi-GBM light curve 8--1000 keV, made with 1-s time bins.  {\sl Bottom:} Swift-BAT light curve, also with 1-s time bins.
The bold dashed lines correspond to a frequency of $\sim$0.04 Hz, and the thin dotted lines correspond to $\sim$0.08 Hz.
}
\label{fig:tdrss_BAT_LC}
\end{figure}

Swift/BAT is a coded-mask instrument that allows arcminute-scale localizations of hard X-ray sources while maintaining a large FoV of $\sim 2$ sr \citep[$> 10\%$ coded;][]{2005SSRv..120..143B}.
The BAT light curve showed a complex structure similar to that detected with GBM. The $T_{90}$ measured in BAT ($15-350$ keV) is $173.43 \pm 5.19$ sec. 
Both the BAT light curve (Fig.\ \ref{fig:tdrss_BAT_LC}, bottom panel) and the GBM light curve (top panel) can be described as consisting of three distinct phases:

\begin{enumerate}
\item the prompt emission, which ranges from $T_0$ to about $T_0+40$ s (post trigger), 

\item a period of quiescence from about $T_0+40$~s to $T_0+105$~s, during which the BAT and GBM count rates return to pre-trigger levels, and  

\item a ``giant flare'' from about $T_0+105$~s to $T_0+175$~s. 
\end{enumerate}

The peak count rate of the prompt emission was $\sim$3,400 net counts~s$^{-1}$ at 12~s post-trigger, and the peak count rate of the giant flare was $\sim$12,000 net counts~s$^{-1}$ ($15-350$ keV), at 162~s post-trigger. The spectrum from $T_0-1.1024$ s to $T_0+225.280$ s is best-fitted by a simple power-law model \citep[see the detailed model description of the model in][] {2016ApJ...829....7L} with a photon index of $-1.64 \pm 0.03$. The fluence in the $15-350$~keV energy range in this time period is $2.64 \times 10^{-5} \, \rm erg \, cm^{-2}$.

\subsection{Swift/XRT}
\begin{figure}[ht]
\centering
\includegraphics[width=0.47\textwidth]{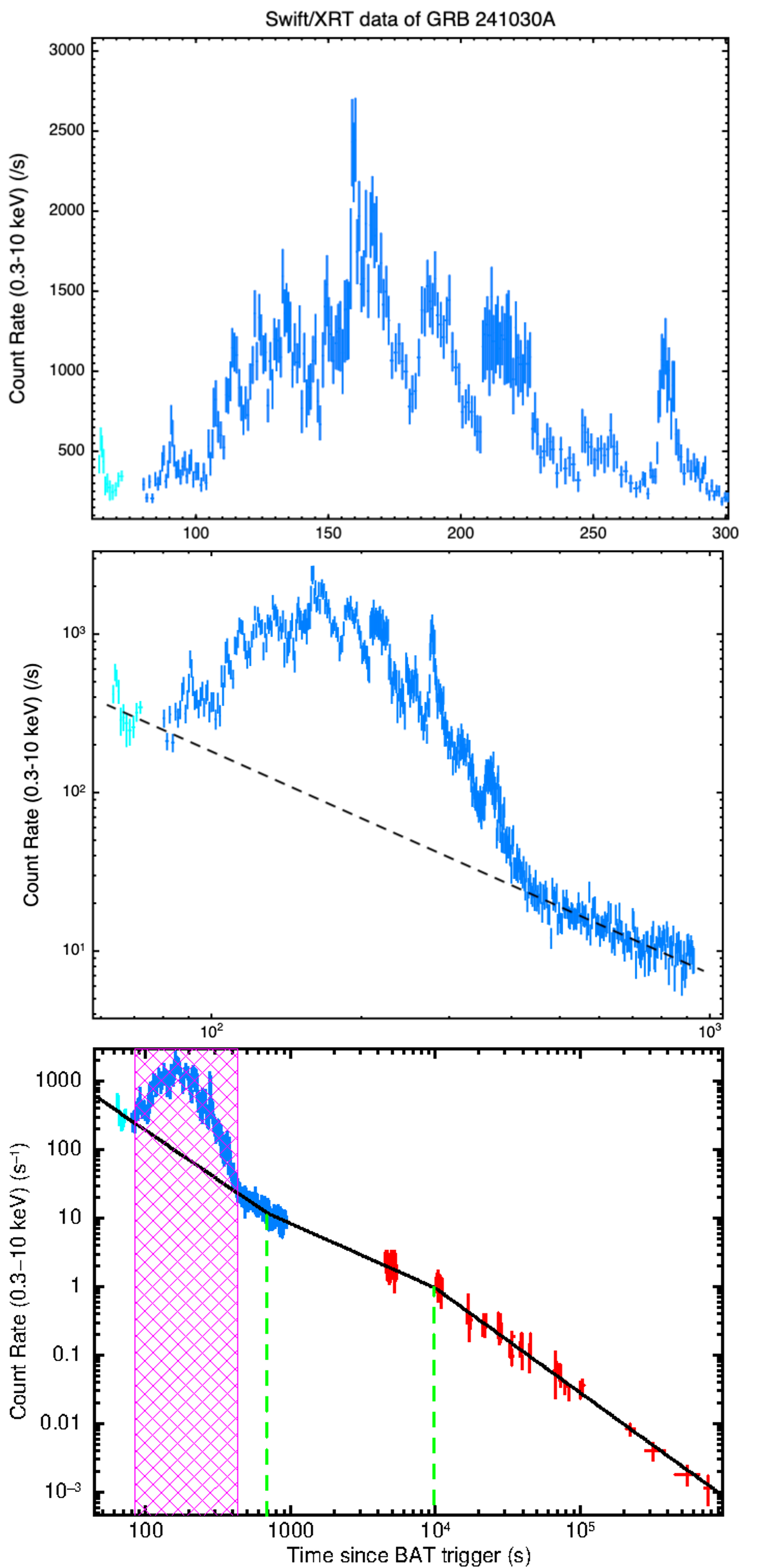}  
\caption{{\sl Top}:  Swift-XRT light curve (linear scales) of the first 300 seconds of the burst, showing the X-ray variability.
{\sl Middle}:  The light curve (in log-log scales) zoomed out to show the entire giant X-ray flare.  
{\sl Bottom}: The entire Swift-XRT light curve. 
The cyan points represent when the telescope was operating in Windowed Timing (WT) mode but still settling on the GRB's location, blue data points are WT data (after settling), and red data points are from Photon Counting (PC) mode. 
The giant flare is highlighted in magenta in the bottom panel.  
The green lines mark the times of the two statistically significant breaks in the light curve.
}
\label{fig:XRT_LCs}
\end{figure}

Due to Swift's rapid slewing capabilities, the XRT's field of view reached the afterglow position at about 60~s post-trigger.
From $60-80$~s, the X-ray emission appears to be fading, but then rises around 80~s, producing a large flaring episode which lasts until about 430~s post-trigger.
The giant flare resembles a Gaussian-like baseline with bright short-term spikes of emission.  
As mentioned above, we refer to these temporal periods as ``prompt emission'' and ``giant flare'' rather than ``precursor'' and ``prompt emission'' because the fading of the X-ray light curve (shown in Figure \ref{fig:XRT_LCs}) appears to consist of a flare event superimposed over the typical power-law (PL) fading which follows prompt emission. 
That is, the X-ray afterglow appears to follow a PL decay before the flare, and returns to a connected PL slope decay after the flare, suggesting that the initial event whose fading trend it follows is the prompt emission (see Figure \ref{fig:XRT_LCs}). 
In X-rays, the giant flare lasts from about 80 to 430~s post-trigger, as was flagged by the UK Swift Science Data Centre's flare filtering algorithm (described in Section 2.4 of \citealt{2009MNRAS.397.1177E}).  

After the flare period, the X-ray emission follows typical GRB fading behavior (count rate $\propto t^{-\alpha}$) which is best described by a PL with 2 breaks:
$\alpha_1 = 1.44 \pm 0.03$,
$T_{\rm break,1} = 689^{+69}_{-75}$~s,
$\alpha_2 = 0.94^{+0.03}_{-0.04}$,
$T_{\rm break,2} = 1010^{+134}_{-185}$~s,
and $\alpha_3 = 1.52\pm0.05$.
The afterglow remained detectable with the XRT (i.e., a count rate $\gtrsim 1 \times10^{-3}$ ct/s) for about 9.26~days post-trigger.

The X-ray and soft gamma-ray spectral properties during the early time emission have been analyzed and reported by \cite{2025ApJ...987..129W}.  
They found that the spectra during the prompt emission and the giant flare (which they refer to as precursor and main prompt emission) cannot be adequately described by a single absorbed power-law (PL) model. 
Instead, joint fits to the XRT and BAT data require the addition of a thermal (blackbody) component, particularly during the giant flare ($\sim 80–430$~s).
The inferred blackbody temperature $kT$ evolves with time, varying between $\approx 1-3$~keV, and generally decreasing as the burst progresses. 
The nonthermal component is well described by a PL with a photon index typical of prompt GRB emission (it varies from $1-2$). 
The inclusion of the thermal component improves the fits significantly during the early-time intervals, indicating that it is statistically required.
This thermal component is interpreted as photospheric emission from the relativistic outflow. 
From the observed temperature and flux, \cite{2025ApJ...987..129W} infer a bulk Lorentz factor of $\Gamma \sim 20 – 80$. 
The presence of thermal and nonthermal components suggests that both photospheric emission and internal dissipation processes (e.g., internal shocks) contribute to the early X-ray emission.
Outside of the flare interval, the XRT spectra are adequately described by a single absorbed PL with no statistically significant thermal component, consistent with a transition to standard afterglow emission dominated by synchrotron radiation from the external forward shock.

\subsection{Swift/UVOT}
\begin{figure}[ht]
\centering
\includegraphics[width=0.5\textwidth]{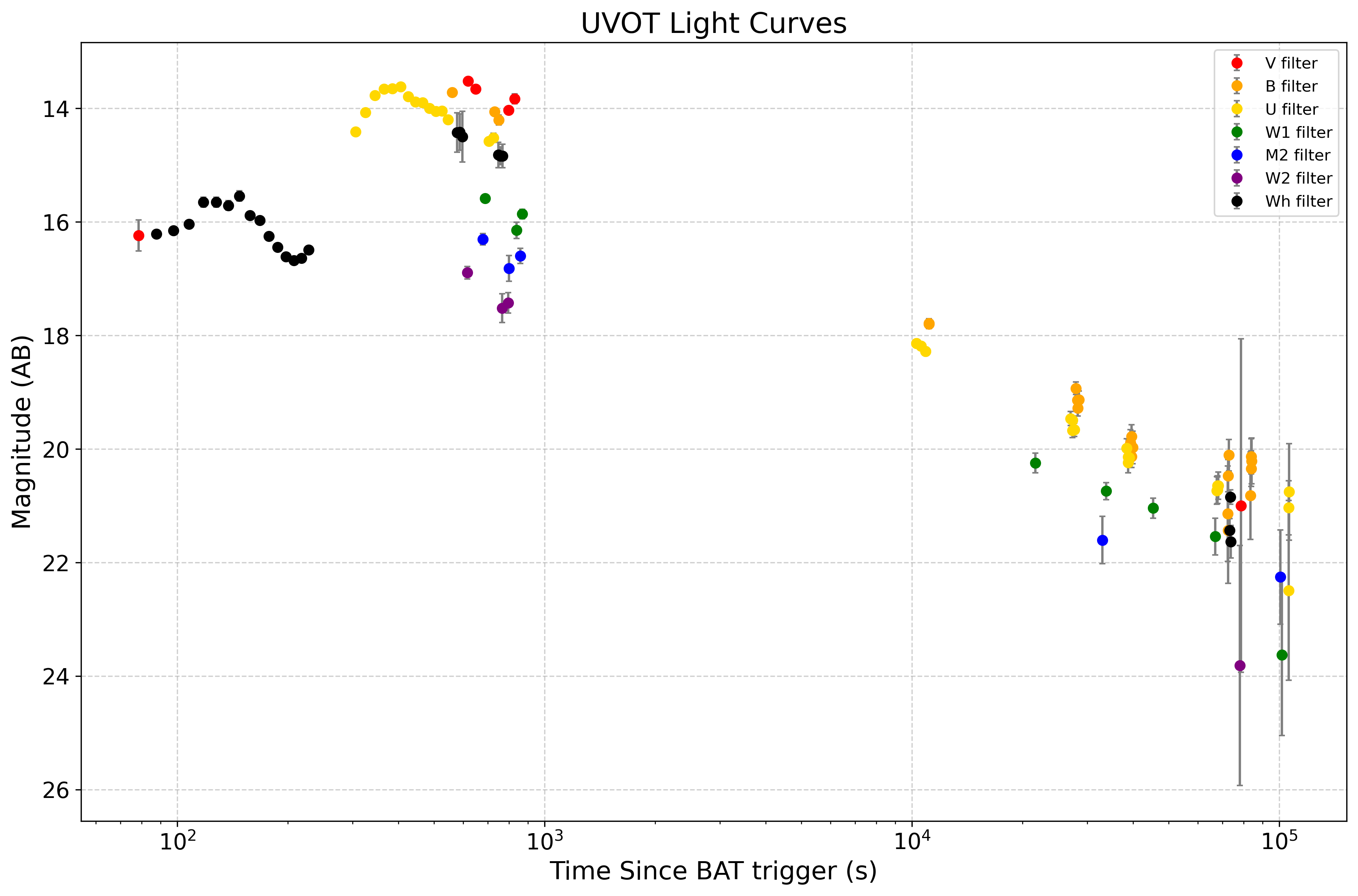}
\includegraphics[width=0.5\textwidth]{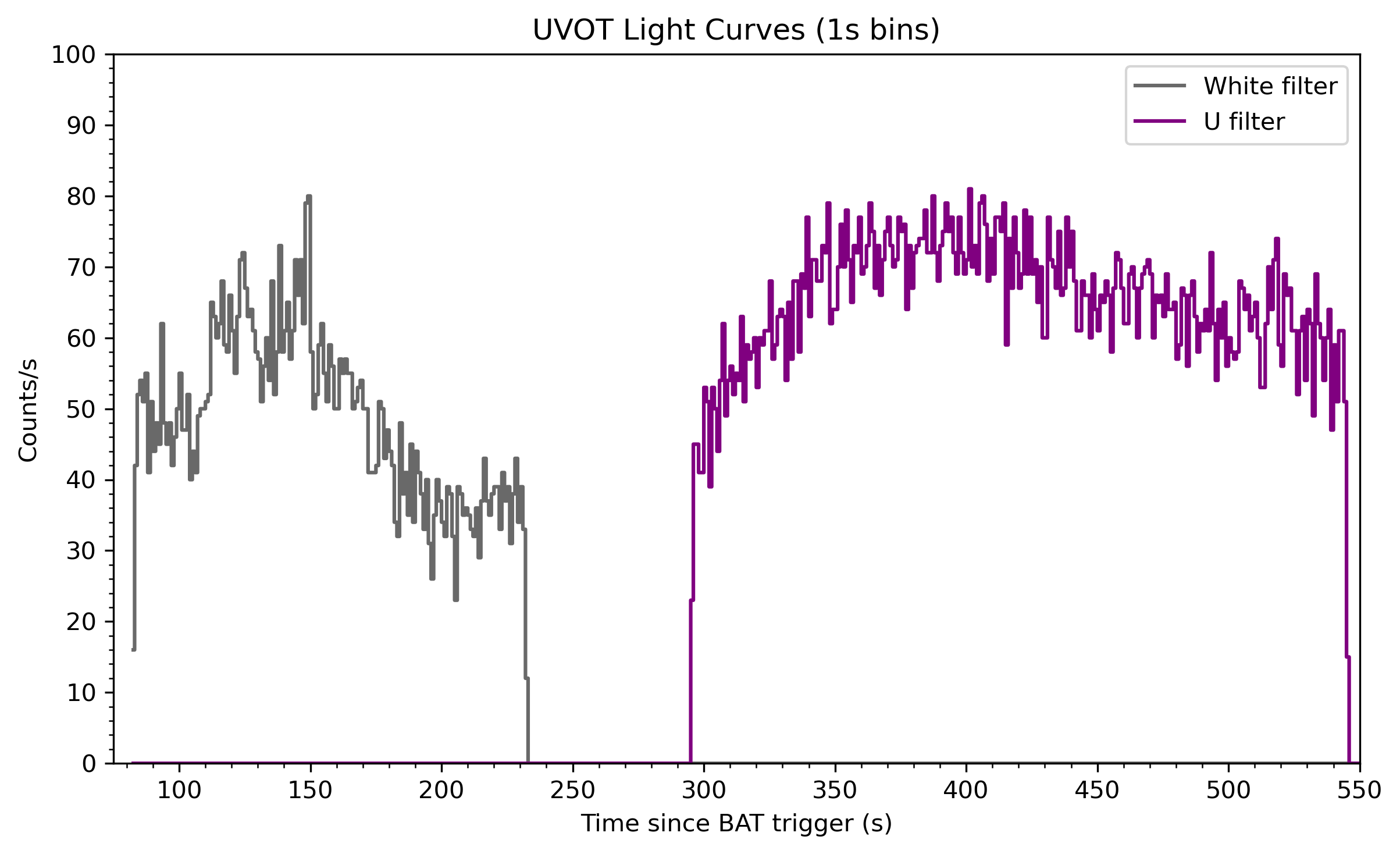}
\caption{
{\sl Top:} UVOT light curves showing all UVOT data / all filters.
{\sl Bottom:} UVOT light curves of the early time emission, made from the event data for 1-second time bins.
The gap in data between the White and U-band coverage was caused by UVOT switching to the UV grism.
}
\label{fig:UVOT_LCs}
\end{figure}

UVOT began collecting data at about 80~s post-trigger, beginning with the White filter.
The burst count rate rises until about 120~s, peaking with a magnitude of about 15.6 (AB), before starting to fade. 
However, around 200~s post-trigger the emission starts to rise again.
The emission then peaks in the U-band at 13.6 mag (AB) at around 400~s, after which the burst begins to fade with a typical PL decay.
The UVOT light curves are shown in Figure \ref{fig:UVOT_LCs}. 
The peak at 400~s can be attributed to the onset of the forward shock emission \citep{2025ApJ...987..129W}. 

The 1-sec binned light curve shows rapid variability in the early time emission (i.e., from 80 - 150 s), somewhat tracing the rapid gamma-ray and X-ray variability.  
Some emission spikes appear to be coincident with those in gamma-rays (e.g., notably the spike around $T_0 + 150$~s). 
The second peak, although brighter, appears to be smoother and less variable.
The UV, optical, and IR properties of this burst have been studied in detail by \citet{2026arXiv260318956D}.

\section{Data Analysis}
\label{sec:gamma}

\subsection{Data}

\subsubsection{Fermi/GBM}
\label{sec:data_gbm}
We used the NaI detectors n0 and n1 for our QPO search in the GBM data, as the signal is brightest in these detectors, in order to enable the most sensitive study of the timing properties.
We used all events reported in the TTE files available in the online repository\footnote{\url{https://heasarc.gsfc.nasa.gov/FTP/fermi/data/gbm/bursts/2024/bn241030242/current/}}, without applying any energy selection (i.e., including all 128 energy channels). 
We used \texttt{HEASoft} package (v6.29) and the \texttt{fdump} tool to extract the data.
The light curve was produced using a time bin of 100-$\mu$s and the power density spectrum was produced using an interval of 2$^{21}$ $\times$ 10$^{-4}$ s = 209.7152 s. 
This choice (where the interval correspond to a power-of-two multiple of the time bin) is used to construct the fast Fourier transform as described in \citet{2023Natur.613..253C}.
\smallskip

\subsubsection{Swift/BAT}
\label{sec:data_bat}
For our QPO analysis we use non-maskweighted (i.e., not background-subtracted) BAT event data from a duration of $T_{\rm 100} \pm 10 \, \rm s$ (i.e., 100\% of the GRB's duration, with the preceding and subsequent 10~s). 
Event data record the energy and timing information for each photon. 
We utilized the non-maskweighted event data because the QPO analysis requires that the data follow Poisson statistics and do not have negative and/or fractional photon counts, and the mask weighting process results in negative and fractional photon counts. 
Following the standard BAT-data processing methods (including {\it batgrbproduct} v2.48, {\it batbinevt} v1.48, {\it batdrmgen} v3.6 and Xspec v12.13.1), we excluded detectors that are either disabled or are known to be noisy from the analysis. 
Both the event data and the list of excluded detectors are obtained from the {\it Swift}/BAT GRB Catalog\footnote{\url{https://swift.gsfc.nasa.gov/results/batgrbcat/}} \citep{2016ApJ...829....7L}.

\subsubsection{Swift/XRT}
\label{sec:data_xrt}

For the XRT data, we utilized the HEASARC pipeline-produced level 2 data products; specifically sw01263718000xwtw2po\_cl.evt.gz: the cleaned event list for segment 0 (the first XRT data taken immediately after slewing to the GRB location, once Swift has finished settling). 
The XRT was operating in Windowed Timing mode during this period, which provides a timing resolution of 1.8~ms. 
We extracted all counts within a 1 arcmin radius of the GRB position, utilizing data from 79.813~s post-trigger (the time at which the post-slew settling was complete) to 430~s post-trigger (roughly the end of the giant flare / the period during which X-ray variability is seen; see Figure \ref{fig:XRT_LCs}).

\subsubsection{Swift-UVOT}
\label{sec:data_uvot}

For the UVOT data, we utilized the segment 0 event lists from the White and U filters (sw01263718000uwhw1po\_uf.evt.gz and sw01263718000uuuw1po\_uf.evt.gz).
UVOT was operating in Event Mode during the initial part of the burst, which provides timing resolution of 11~ms. 
For the White and U filters, we extracted all events within a 4$''$ and 6$''$ radius of the GRB's position, respectively, due to the slightly different point spread functions in each band. 
The White filter provided continuous coverage from 82.6 to 232.3~s post-trigger, and the U filter provided continuous coverage from 295.5 to 545.2~s post-trigger.
In the top panel of Figure \ref{fig:UVOT_LCs} we show the long-term UVOT light curve, and in the bottom panel we focus on the first few hundred seconds of the GRB by showing the 1-s binned light curve.
The long-term light curve and evolution of the GRB are discussed in detail by \citet{2026arXiv260318956D}.

\subsection{QPO Searches for GRB 241030A}

To the eye, the light curve of GRB~241030A is highly unusual: as seen in both panels of Figure~\ref{fig:tdrss_BAT_LC}, the burst appears to have $\sim 10$ or more peaks that are in phase with a period of 
$\sim 12-13$~seconds.
In this section we detail our search for QPOs in this burst, which appear to detect a significant QPO at that period.  
However, as we discuss, the substantial and complex red noise for this and similar bursts renders it difficult to distinguish cleanly real QPOs from the happenstance of low-frequency noise.

\subsubsection{$Z^2_1$ Test}

We first attempted to search for QPOs using the $Z^2_1$ statistic (Rayleigh test), which is appropriate for event-mode data and is equivalent to the Fourier power at a given frequency for unbinned photon arrival times \citep{1983A&A...128..245B}. 
We constructed a periodogram sampled at the Fourier resolution, $\Delta f = 1 / T_{\rm obs}$, where $T_{\rm obs}$ is the total duration of the light curve segment analyzed. 
This is the standard frequency spacing corresponding to independent Fourier modes for a uniformly sampled time interval of duration $T_{\rm obs}$.

Under the null hypothesis of no coherent signal, the $Z^2_1$ periodogram values at independent Fourier frequencies are exponentially distributed about the underlying power density spectrum (PDS).
Therefore we modeled the PDS with a parametric function and fitted it using a likelihood appropriate for power spectral estimates that follow a $\chi^2$ distribution with 2 degrees of freedom about the true spectrum (e.g., \citealt{2005A&A...431..391V,2026AJ....171..124E}). 
We utilized a bent power-law model with an additive constant noise floor (Equation 2 in \citealt{2016A&A...589A..98G}, which was found to fit the PDSs of many GRBs).

After fitting the PSD continuum, we assessed candidate QPO features by comparing the observed $Z^2_1$ values to the local statistical distribution implied by the best-fit continuum model.
Because $Z^2_1$ values are exponentially distributed about the continuum level, a local single-trial significance threshold was computed relative to the modeled continuum at each frequency. 
We then evaluated the 3$\sigma$ local threshold corresponding to the appropriate tail probability of the exponential distribution.

\begin{figure}[ht]
\centering
\includegraphics[width=0.48\textwidth]{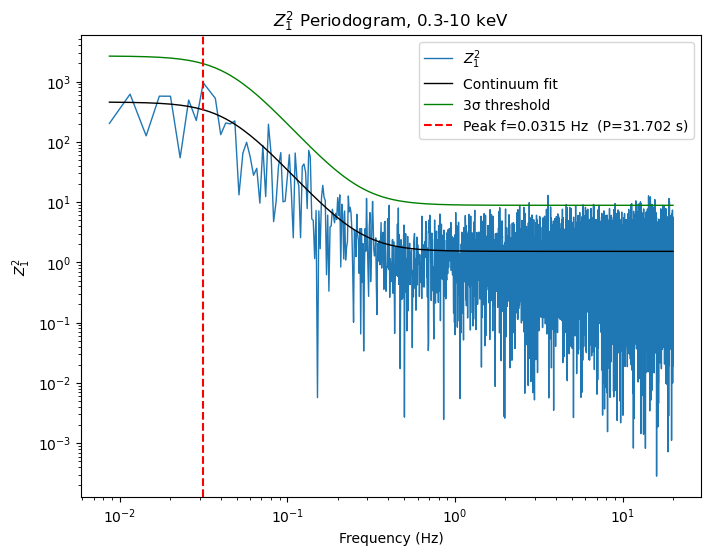}
\caption{$Z^2_1$ periodogram for the Swift-XRT data (0.3--10 keV).  The plot indicates that the putative frequency $\sim0.03 - 0.04$~Hz lies in a frequency range contaminated with red noise, and that the $Z^2_1$ QPOs are not statistically significant when utilizing the $Z^2_1$ test.
The plots for the other energy ranges (0.3--2 keV, 2--10 keV) look virtually the same.
}
\label{fig:Z2_periodograms_X-rays}
\end{figure}

We calculated the $Z^2_1$ values between frequencies 0.008695~Hz ($\approx 1/(3T_{\rm obs})$) and 20~Hz (which is far below the Nyquist frequency $1/(2\Delta t) \approx 278$~Hz).  
We searched the GBM, BAT, XRT, and UVOT data. 
For the X-ray data we applied this procedure for three different X-ray energy ranges: 
$0.3-10$~keV ($124\,786$ counts), 
$0.2-2$~keV ($56\,034$ counts), 
and $2-10$~keV ($69\,093$ counts) 
in order to separate the thermal component found in the X-ray spectra by \citet{2025ApJ...987..129W}.
There are no statistically significant peaks in any of the instruments' data. 
We show the result for the XRT data in Figure \ref{fig:Z2_periodograms_X-rays}; 
the other plots (for GBM, BAT, and UVOT) look qualitatively the same. 
The power peaks in the 0.03--0.04~Hz range, although it is not formally significant. 
A few peaks marginally surpass the 3$\sigma$ threshold at higher frequencies, but these represent statistical occurrences due to the number of frequencies searched (i.e., high number of trials), rather than significant results.

Formally, under this single-trial criterion, we can not claim a statistically significant QPO detection using the $Z^2$ test. 
However, the putative periodicity (e.g., hinted at by eye) lies at low frequencies $\sim 0.03-0.04$ Hz, where the variability is dominated by red noise.  
In this regime, the continuum power rises steeply toward low frequencies, and small deviations from the assumed continuum model can strongly affect local significance estimates. 
The standard thresholding approach of testing individual frequencies  against an exponential null distribution about a fitted continuum may therefore lack sensitivity to broad or weak quasi-periodic features embedded in the red noise.
QPOs can have contributions from more than one frequency, but the standard $Z^2_1$ analysis is  not sensitive to this.  
Because the $Z^2_1$ test evaluates each frequency independently, it does not explicitly compare evidence for competing models of the power spectrum. 
In particular, it does not quantify whether a model including a QPO component provides a statistically meaningful improvement over a red-noise-only description of the data.

\subsubsection{Model Comparison}
\label{sec:data_bayes}

\begin{figure*}[tbh]
\centering
\includegraphics[width=0.45\textwidth]{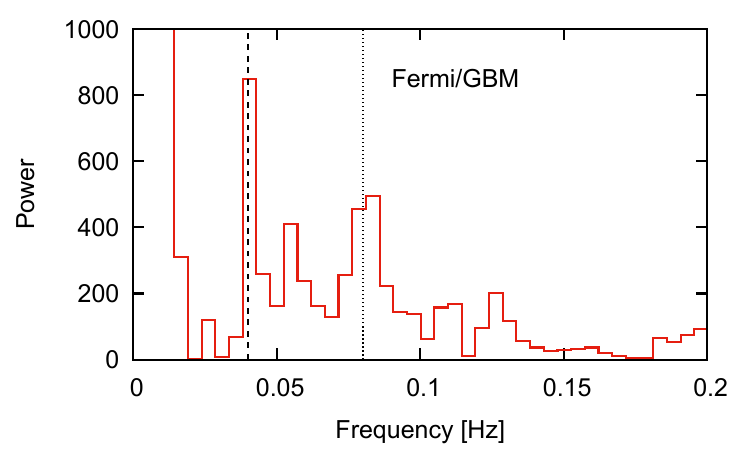}
\includegraphics[width=0.45\textwidth]{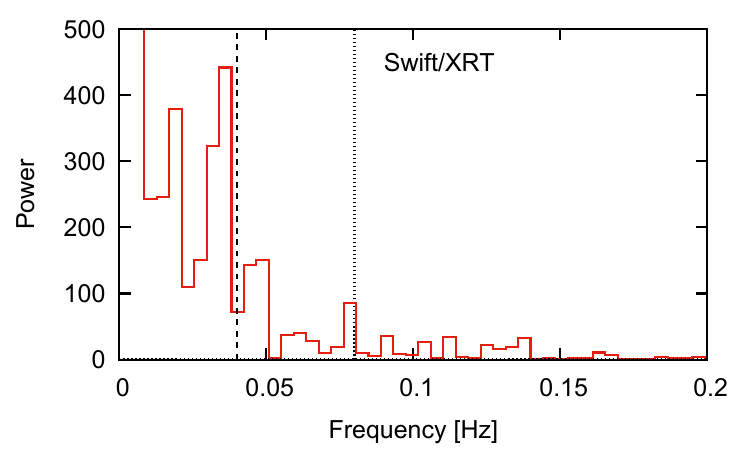}
\includegraphics[width=0.45\textwidth]{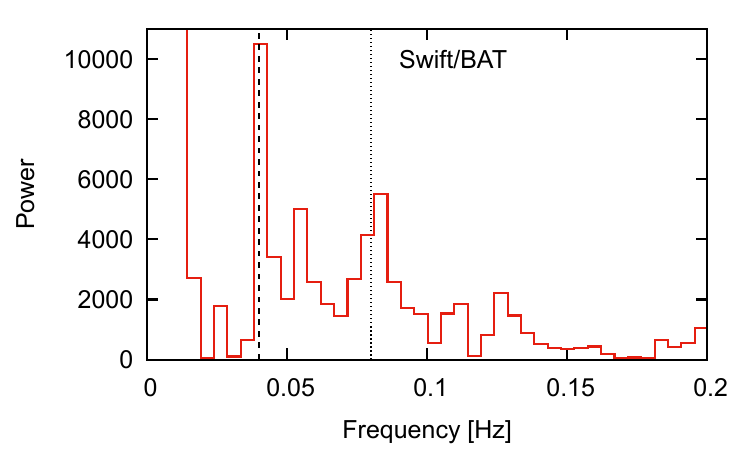}
\includegraphics[width=0.45\textwidth]{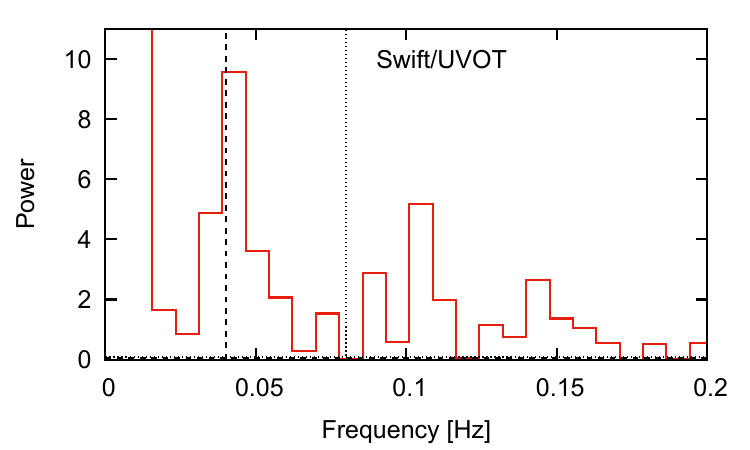}
\caption{Power spectra of the GBM, BAT, XRT and UVOT light curves shown in Figures \ref{fig:tdrss_BAT_LC}--\ref{fig:UVOT_LCs}, normalized so that the probability of a power $>P$ from pure Poisson noise is $e^{-P}$. 
Power rising sharply with decreasing frequency for $f \lesssim 0.02$ Hz indicates strong red noise; the highlighted peaks in the spectrum are consistent with a ``fundamental frequency'' $f_0 \simeq 0.04$ Hz (vertical dashed line) and a ``first harmonic'' at $2f_0$ (vertical dotted line). 
The QPO at $f_0$ is statistically significant in the GBM, BAT and XRT data, but not in the UVOT data, although the UVOT data also show a clear power excess at $\sim f_0$. 
See more details in Sec. \ref{sec:results}. 
Note the trend with energy, in that the $2f_0$ QPO is not significant in X-rays, and also note that there is excess power at $\sim 3f_0$ seen in the BAT and GBM data.
}
\label{fig:spectrum}
\end{figure*}

Our search for QPOs in GRB~241030A is made more complicated because the periods apparently evident to the eye in the power spectra are low enough ($\sim 0.04$~Hz and $\sim 0.08$~Hz) that there is substantial red noise competing with the QPO (see Figure~\ref{fig:spectrum}).  
We therefore need to perform a model comparison between a power spectrum with just red noise, and one which in addition has a QPO.

To do this, we use the approach first developed by \citet{2019ApJ...871...95M} in the study of QPOs in giant flares from soft gamma-ray repeaters, which was then applied to gamma-ray bursts in \citet{2023Natur.613..253C} and \citet{2024ApJ...967...26C}.  This is a Bayesian model comparison method, applied to the power spectral data.  One model has just red noise, whereas the other has red noise plus a Lorentzian which represents the QPO.  As usual, the Bayes factor between the two models is the ratio of their evidences:
\begin{equation}
    {\cal B}_{\rm QPO,red}\equiv\frac{\int {\cal L}({\rm data}|{\vec\alpha}_{\rm QPO})q({\vec\alpha}_{\rm QPO})d{\vec\alpha}_{\rm QPO}}{\int {\cal L}({\rm data}|{\vec\alpha}_{\rm red})q({\vec\alpha}_{\rm red})d{\vec\alpha}_{\rm red}}\; .
\end{equation}
Here ${\cal L}$ is the likelihood of the data given the model (see below), $q$ is the normalized prior, ${\vec\alpha}_{\rm QPO}$ represents the parameters in the model with a QPO and red noise, and ${\vec\alpha}_{\rm red}$ represents the parameters in the model with just red noise.

\begin{table}
    \centering
    \caption{Priors for our QPO analyses}
    \label{tab:priors}
    \begin{tabular}{ll}
    \hline
    Parameter & Prior\\
    \hline
    Amplitude at lowest freq.\ $f_{\rm low}$ & Uniform, 0 to $10^7$ \\
    \hline
    Power law $f^{-\alpha_2}$ to 2nd freq.\ & Uniform, $\alpha_2=-1$ to $+3$\\
    \hline
    Power law $f^{-\alpha_3}$ to 3rd freq.\ & Uniform, $\alpha_3=-1$ to $+3$\\
    \hline
    Power law $f^{-\alpha_4}$ to 4th freq.\ & Uniform, $\alpha_4=-1$ to $+3$\\
    \hline
    Power law $f^{-\alpha_5}$ to $f=\infty$ & Uniform, $\alpha_5=-1$ to $+3$\\
    \hline
    Lorentzian peak power & Uniform, 0 to $10^6$\\
    \hline
    Lorentzian centroid $f_c$ & Log unif, $0.02-0.1$~Hz\\
    \hline
    Lorentzian width $\Delta f$ & Log unif, $10^{-4}-10^{-1.5}$~Hz$^{\rm a}$\\
    \hline
    \end{tabular}
\begin{flushleft} 
\quad \footnotesize{$^{a}$ In Sec. \ref{sec:results_pop}, we impose the additional requirement that} $\Delta f\geq f_{\rm low}$ and $f_c/\Delta f\geq 10$.\\ 
\end{flushleft}
\end{table}

A question in such comparisons is how to represent the red noise.  Previous analyses, particularly of the ``long short burst" GRB~211211A \citep{2024ApJ...967...26C}, show that a single power law model for the red noise is insufficiently flexible.  We therefore follow \citet{2024ApJ...967...26C} in using a model where the lowest four frequencies are connected by power laws of possibly different slopes, and the higher frequencies are connected with a single red-noise power law.  The model with a QPO has the same red noise description as the model with just red noise, and in addition has a Lorentzian which describes the QPO.  See Table~\ref{tab:priors} for our priors in these analyses.

To perform our analysis we then need to define a likelihood function.  Following \citet{2019ApJ...871...95M} we use the power spectrum normalization and probability distributions of \citet{1975ApJS...29..285G}.  In this normalization, pure Poisson fluctuations around an intrinsically constant flux produce a power spectrum with an average power of 1. Using this normalization, \citet{1975ApJS...29..285G} found that, for a single frequency bin, the likelihood that a power $P$ would be measured if the source power is $P_s$ (where $P_s=0$ means pure Poisson fluctuations without intrinsic variation) is
\begin{equation}
    {\cal L}(P|P_s)=e^{-(P+P_s)}\sum_{m=0}^\infty P^mP_s^m/(m!)^2\; .
\end{equation}
In the special case $P_s=0$ (pure Poisson noise) the probability is $e^{-P}$, but in general one must calculate the sum.

The likelihood of the power spectrum given a model (either just red noise or red noise plus a QPO) is then the product of these likelihoods over all frequencies in the power spectrum.  
We compute power spectra from data sets with a power of two number of frequencies, and then perform a fast Fourier transform, so that the frequency bins are formally independent. 

However, we echo \citet{2022ApJ...936...17H} in cautioning that when the overall light curve changes rapidly (as it does in GRBs), this can effectively correlate different frequencies in a power spectrum.  
Thus, although our Bayes factors are enormously in favor of the QPO model, it is necessary to compare with other light curves, synthetic or real, to judge whether this constitutes definitive evidence for a QPO. 

\nk{}

\subsection{Population Analysis}
\label{sec:population}

In order to compare the results obtained for GRB 241030A with the population of GRBs detected by \textit{Fermi}/GBM, we retrieved the time-tagged event (TTE) files for all GRBs in the GBM online catalog \citep{2020ApJ...893...46V}\footnote{\url{https://heasarc.gsfc.nasa.gov/w3browse/fermi/fermigbrst.html}} with T$_{90}$ $>$ 2 seconds, discovered up to April 21, 2024. The total number of events considered is 3315. Out of those, we kept the roughly 90 GRBs which, like GRB~241030A, have durations between $2^{20.5}\times 10^{-4}\approx 148~{\rm s}$ and $2^{21.5}\times 10^{-4}\approx 297~{\rm s}$, where as before $10^{-4}$ seconds is the time resolution that we use for the Fermi data. We then rebin each data set so that it fits exactly in $2^{21}=2,097,152$ equal-length intervals. This allows us to use an FFT for our analysis, and keeps the number of independent frequencies constant for all GRBs in the set. For each source, we selected the set of NaI detectors that triggered GBM and were used for the calculation of the GRB duration. These detectors were identified using the ``Bcat\_Detector\_Mask'' catalog entry. For our analysis, we used all photons collected during the T$_{90}$ time interval (as reported in the catalog) and combined them to achieve a higher signal-to-noise ratio. 
The data were downloaded from the GBM online repository\footnote{\url{https://heasarc.gsfc.nasa.gov/FTP/fermi/data/gbm/bursts/}}. 
Seven GRBs (out of the 3315) were excluded from the analysis due to the unavailability of TTE data. The trigger names of the excluded events are: bn101214993, bn101220864, bn110428388, bn110618760, bn110828575, bn121123421, and bn120702891.
We used the \texttt{fdump} tool to extract information for each photon detected by the instrument, and then reconstructed the light curves using a 100-$\mu$s time bin.
We then performed the model comparison method presented in the previous Section to derive the Bayes factors.

In order to assess the time evolution of the QPOs we also computed power spectra in sliding windows.  Each window has half the length of the full data set.  We increment the starting time of the window by intervals of 1/16 of the data set; thus the first window goes from the beginning of the data set to halfway through the data set, the second starts at 1/16 of the way through the data set to 9/16 of the way through the data set, and so on to the ninth, which starts at 1/2 of the way through the data set and finishes at the end of the data set.

\section{Results}
\label{sec:results}

\subsection{QPO Search for GRB 241030A}
\label{sec:results1}

In Figure \ref{fig:spectrum} we show the power spectrum of the light curves of GRB 241030A obtained with different instruments in different wavelengths:  gamma-rays (Fermi/GBM and Swift/BAT), X-rays (Swift/XRT) and UV (Swift/UVOT; specifically the $U$-band). 
The GBM and BAT data cover the same duration, while the XRT and UVOT data start later due to the slew time (see Secs.\ \ref{sec:data_gbm}-\ref{sec:data_uvot} for more details). 
We used the data in the frequency domain for our analysis, as described in Sec.\ \ref{sec:data_bayes}.
For the BAT data, we verified that the nearby bright variable sources which were also in the FOV (Cygnus X-1, X-2, and X-3) did not meaningfully affect our results (see Appendix Sect.\ \ref{sec:appendix1} for details).

GRB 241030A has a statistically significant\footnote{In this context, ``significant'' means that the Bayes factor is enormous and favors the QPO + red noise model over the red-noise-only model.} QPO at $f_0 \approx 0.04$ Hz, corresponding to a period $P \approx 25$ s, in the GBM, BAT, and XRT data (see Table~\ref{tab:best0.04} for the best-fit parameters). 
The UVOT data also have excess power at this frequency, but the power excess is much lower than in the other wavelengths (see Fig.\  \ref{fig:spectrum}). 
We also see an excess of power at frequencies consistent with harmonics: $2f_0$ in XRT, BAT, and GBM data; and $3f_0$ in BAT and GBM data.
To the best of our knowledge, this is the first reported GRB QPO  detected at a consistent frequency in both soft X-rays (Swift/XRT) and in gamma-rays (Swift/BAT and Fermi/GBM).

\begin{table*}[th]
    \centering
    \caption{Best fits for $\sim 0.04$~Hz QPO}
    \label{tab:best0.04}
    \begin{tabular}{lllllllllc}
    \hline
    Data Set & $A(f_{\rm low})$ & $\alpha_2$ & $\alpha_3$ & $\alpha_4$ & $\alpha_5$ & $P_{\rm peak}$ & $f_c$ (Hz) & $\Delta f$ (Hz) & $\log_{10}\mathcal{B}_{\rm{QPO},\rm{red}}^{\rm a}$\\
    \hline
    GBM & 1977 & 0.94 & 3.0 & 3.0 & 0.90 & 400 & 0.038 & 0.0001 & $\sim 182$\\
    \hline
    BAT & 56026 & 1.26 & 3.0 & 2.83 & 0.97 & 19998 & 0.039 & 0.001 & $\sim 50600$ \\
    \hline
    XRT & 49750 & 3.0 & 3.0 & 3.0 & 1.94 & 401 & 0.033 & 0.003 & $\sim 150$\\
    \hline
    UVOT & 638 & 2.57 & 0.81 & 2.99 & 2.62 & 24 & 0.040 & 0.001 & -2.7$^{\rm b}$\\
    \hline
    \end{tabular}
\begin{flushleft} 
\quad \footnotesize{$^{a}$ The large Bayes factors here and in Table~\ref{tab:best0.08} are approximate, given the sensitivity to exact log likelihood values.}\\
\quad \footnotesize{$^{b}$ The UVOT data show frequency peaks consistent with the other bands, but the data do not have enough counts to claim evidence for the QPOs from the UVOT data alone.\\
\quad Note -- several of the $\alpha$ values are pegged to their prior maximum, 3.0 (see Table~\ref{tab:priors}).}
\\ 
\end{flushleft}
\end{table*}

\begin{figure}[tbh]
\centering
\includegraphics[width=0.45\textwidth]{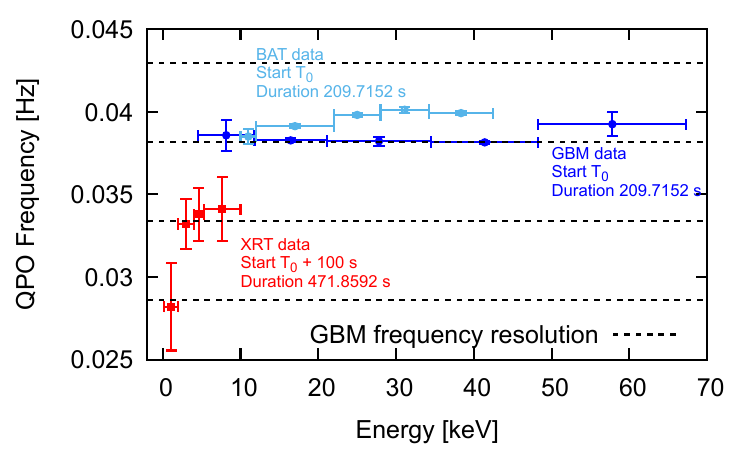}
\caption{
We find no indication of frequency evolution with time or energy in the GBM data. The horizontal error bars indicate the independent energy ranges that were analyzed separately (keeping the number of counts approximately constant in all energy ranges for each detector); the vertical error bars indicate the width of the QPO, but they do not fully incorporate modelling systematics, e.g. the QPO may not be perfectly described by a Lorentzian. Note that the XRT data and GBM/BAT data only have partial time overlap (which may be responsible for the slight discrepancy in the peak QPO frequency in the range of $\lesssim 10$ keV). The XRT and GBM QPO frequencies are consistent in the same energy range within the GBM frequency resolution of $1/(209.7152~{\rm s})$, indicated by dashed horizontal lines.}
\label{fig:GBM+XRT}
\end{figure}

In Fig.~\ref{fig:GBM+XRT} we explore the energy dependence of the QPO frequency $f_0$. 
The QPO frequencies obtained with the GBM and BAT data are approximately constant in this energy range, without an obvious trend. The vertical error bars underestimate the uncertainty in the QPO frequency (see Fig.\ \ref{fig:GBM+XRT} caption for more details), causing an apparent inconsistency between the GBM frequencies and the BAT frequencies. 
The XRT data show a clearer frequency evolution with energy; it is roughly consistent with the GBM data (in their overlapping energy range) within the GBM frequency resolution, but note that both data sets correspond to different time intervals, with only a partial temporal overlap.

\subsection{Population Analysis}
\label{sec:results_pop}

As shown in Sec.\ \ref{sec:results1}, GRB 241030A has a very significant QPO (in terms of the Bayes factor $\mathcal{B}_{\rm{QPO},\rm{red}}$) at the frequency $f_0 \sim 0.04$ Hz. 
A separate question that needs to be addressed is whether such a strong QPO could be mimicked by red noise in a GRB light curve with no quasi-periodicity. 
The difficulty of distinguishing real periodicity from mimicked periodicity is further highlighted by recent work showing that QPO-like light curves can be reproduced by a stochastic pulse avalanche process \citet{2024A&A...689A.266B}. 
In such models, a prompt emission pulse can stochastically trigger subsequent pulses, producing an avalanche-like cascade whose superposition generates complex multi-peaked GRB light curves, without actually having an underlying periodic engine.

A standard procedure is to simulate and analyze similar light curves produced without QPOs. However, this procedure is complicated in our case by the complex temporal structure of GRB 241030A, as it is unclear what features a ``similar'' light curve should have. We choose, instead, to analyze the sample of $\sim 90$ long GRBs described in Sec. \ref{sec:population} and use the results as a baseline to compare with GRB 241030A.

\begin{table*}[th]
    \centering
    \caption{Best fits for $\sim 0.08$~Hz QPO}
    \label{tab:best0.08}
    \begin{tabular}{lllllllllc}
    \hline
    Data Set & $A(f_{\rm low})$ & $\alpha_2$ & $\alpha_3$ & $\alpha_4$ & $\alpha_5$ & $P_{\rm peak}$ & $f_c$ (Hz) & $\Delta f$ (Hz)& $\log_{10}\mathcal{B}_{\rm{QPO},\rm{red}}$\\
    \hline
    GBM & 1915 & 0.99 & 3.0 & 3.0 & 0.93 & 218 & 0.080 & 0.007 & $\sim 229$\\
    \hline
    BAT & 55654 & 1.31 & 3.0 & 3.0 & 1.00 & 4658 & 0.079 & 0.008 & $\sim 52700$ \\
    \hline
    XRT & 49896 & 3.0 & 3.0 & 3.0 & 1.54 & 57 & 0.076 & 0.001 & 1.22$^{\rm a}$\\
    \hline
    UVOT & 629 & 2.55 & 0.68 & 2.95 & 2.62 & 7.4 & 0.078 & 0.001 & -2.7$^{\rm b}$\\
    \hline
    \end{tabular}
\begin{flushleft} 
\quad \footnotesize{$^{a}$ The XRT data showed only weak evidence for this QPO.}\\ \quad \footnotesize{$^{b}$ The UVOT data showed no evidence for this QPO.}\\ 
\end{flushleft}
\end{table*}

In order to increase the chances that a detected QPO is real rather than being a happenstance peak of the noise, we supplement the priors listed in Table \ref{tab:priors} with two additional conditions (see footnote in Table \ref{tab:priors}) intended to: (1) avoid a QPO with a best-fit centroid frequency at the lowest frequency in the power spectrum (typically the frequency with the highest power due to red noise); and (2) require that the centroid frequency of the QPO be at least ten times the width of the QPO, to impose a high quality factor for the QPO (which should correspond to more oscillation periods in the light curve). 

\begin{figure*}[tbh]
\centering
\includegraphics[width=0.75\textwidth]{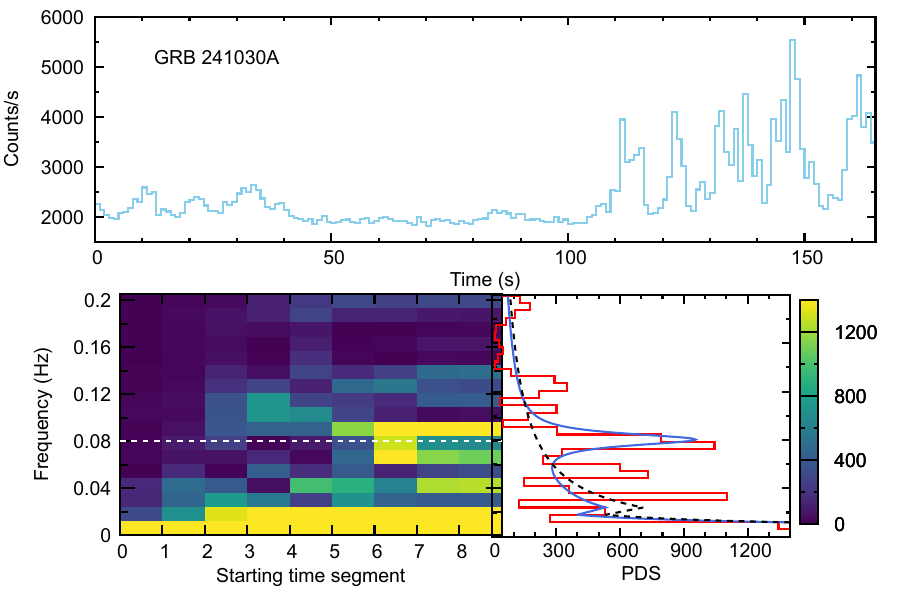}
\caption{Top: Light curve of GRB 241030A, showing only the $T_{90}$ duration of the burst, used for our analysis.  
Thus, here, time 0 corresponds to the start of T90, which is 20.992 s post-trigger. 
Bottom left: Spectrogram of GRB 241030A, obtained using a sliding window with a duration of $T_{90}/2$ and a sliding step of $T_{90}/16$. The horizontal dashed white line shows the best-fit value of the QPO centroid frequency (the frequency values are matched in the y-axes of both bottom  plots). Bottom right: PSD of GRB 241030A overlaid with the best fits for our two models: the dashed black line indicates the best fit for the noise-only model, and the solid blue line shows the best fit for the QPO+noise model.
}
\label{fig:spectrogram}
\end{figure*}

As a result of the addition of these extra conditions in the priors, our analysis of GRB 241030A finds the ``first-overtone'' QPO with approximate frequency $2f_0 \sim 0.08$ Hz (see Fig. \ref{fig:spectrum}). In Fig. \ref{fig:spectrogram} we show the light curve, dynamical power spectrum, and power spectrum with the best noise-only and the best noise+QPO fits for GRB 241030A. 
In Table~\ref{tab:best0.08} we show the best-fit parameters for the $\sim 0.08$~Hz QPO in each of the data sets. 
Comparing Table~\ref{tab:best0.04} and Table~\ref{tab:best0.08} we see that the centroid frequencies are at least consistent with being harmonics of an underlying oscillation.

\begin{table}
    \centering
    \caption{GBM Data for the other 6 GRBs with high Bayes factor in Fig. \ref{fig:rms_bayes}}
    \label{tab:grb_counts}
    \begin{tabular}{lccc}
    \hline
    GRB & Trigger Name & NaI detectors$^{a}$ & Tot. Counts$^{b}$ \\
    \hline
    101014A & bn101014175 & n6-n7-n8-nb & 1266491 \\
    210204A & bn210204270 & n6-n7-n9-nb & 1217094 \\
    210606B & bn210606945 & n3-n4-n5 & 1054036 \\
    211130A & bn211130636 & n6-n7-n8 & 723709 \\
    220921A & bn220921462 & n0-n1-n2-n9-na & 1295067 \\
    250313A & bn250313607 & n0-n1-n3-n6-n7-n9 & 2271897 \\
    \hline
    \hline
    \end{tabular}
\begin{flushleft} 
\quad \footnotesize{$^{a}$ NaI detectors used for the analysis.}\\ 
\quad \footnotesize{$^{b}$ Total number of counts collected by the different detectors during the T$_{90}$ interval.}\\ 
\end{flushleft}
\end{table}

\begin{figure*}[tbh]
\centering
\includegraphics[width=1\textwidth]{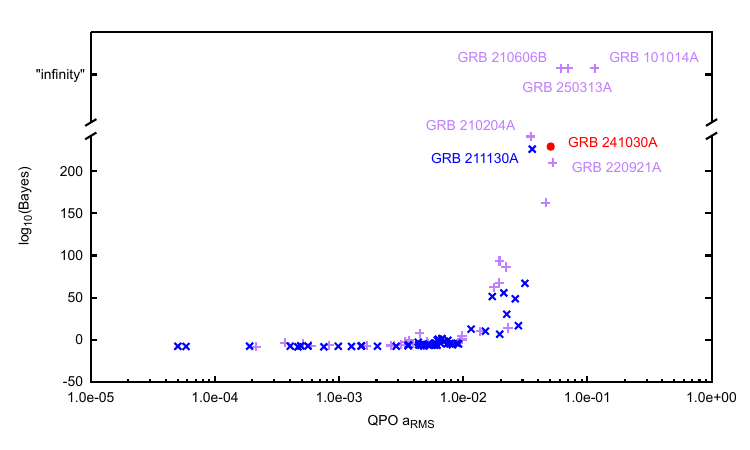}
\caption{Bayes factor vs. fractional RMS amplitude $a_{\rm RMS} = \sqrt{2P_{\rm QPO}/N_{\rm Total}}$ for the sample of $\sim 90$ long GRBs. The blue points ($\times$) correspond to bursts with $N_{\rm total} < 10^6$, and the purple points ($+$) correspond to bursts with $N_{\rm total} > 10^6$. Most of the sample has a low Bayes factor in favor of the QPO model, and a corresponding low fractional RMS amplitude for the best-fit QPO frequency identified in each case. GRB 241030A stands out from the blue distribution. The lone blue point closest to it represents GRB 211130A, which we analyze separately in Fig. \ref{fig:fake1}. 
}
\label{fig:rms_bayes}
\end{figure*}

With this set of priors we analyze our full sample of $\sim 90$ long GRBs. 
In Fig. \ref{fig:rms_bayes} we show the 2D scatter plot of Bayes factors (in favor of the QPO model) and the RMS amplitude $a_{\rm RMS} = \sqrt{2P_{\rm QPO}/N_{\rm total}}$ of the best-fit QPO in each case. Most of the sample has a low Bayes factor in favor of the QPO model (approximately 70 GRBs). 
The remaining GRBs have very high Bayes factors and most of them are very bright bursts with very high total number of counts $N_{\rm total} \gtrsim 10^6$ (shown in purple; see Table~\ref{tab:grb_counts}). 
High count numbers could increase the chance that noise fluctuations are mistaken for QPOs, given that power $\propto N_{\rm total}$. 

\begin{figure}[tbh]
\centering
\includegraphics[width=0.45\textwidth]{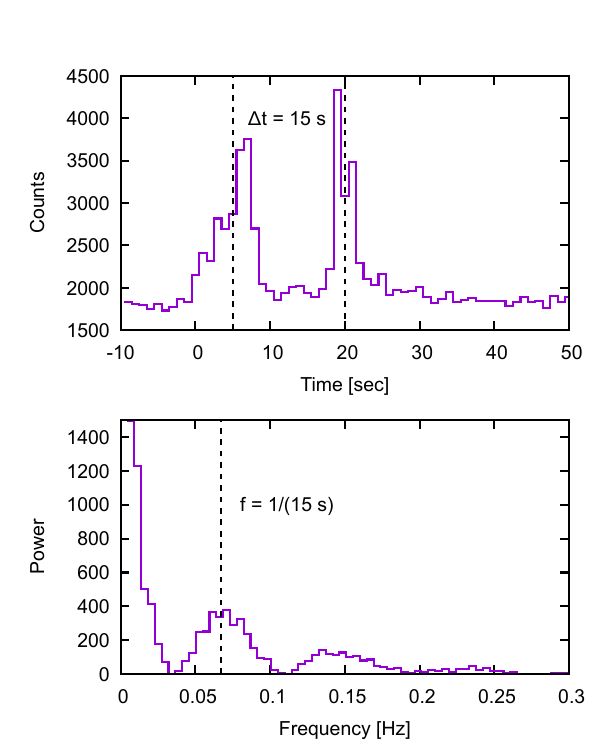}
\caption{Example of a GRB with a fake QPO, GRB 211130A, detected with Fermi/GBM. The large number of counts causes high power in the PSD, and the separation between the two peaks produces a QPO-like feature. This GRB has a Bayes factor larger than $10^{200}$ in favor of the QPO model, even though there is obviously no periodicity/quasiperiodicity in the light curve.}
\label{fig:fake1}
\end{figure}

GRB 241030A stands out from the blue distribution shown in Fig. \ref{fig:rms_bayes}, which comprises the subset of GRBs in our sample that have $N_{\rm total} < 10^6$. 
The only GRB in this sample with a comparable Bayes factor to that of GRB 241030A is GRB 211130A. 
We analyze it separately in Fig.\ \ref{fig:fake1}.  Unlike GRB 241030A, there are no repeating oscillations in the light curve of GRB 211130A. Instead, GRB 211130A has a double-peaked light curve, shown in the top panel of Fig.\ \ref{fig:fake1}, where the time separation between both peaks is approximately $\Delta t = 15$ s. This causes a power excess at a frequency $1/\Delta t$, which is picked up by our algorithm as a QPO candidate. This ``fake'' QPO is further enhanced by (1)~the high number of counts, which drive up the power, and (2)~the long duration of the GRB ($T_{90} > 200$~s) compared with the segment that shows the peaks in the light curve, which over-resolves the power spectrum, see also \citet{2022ApJ...936...17H}.

\begin{figure*}[tbh]
\centering
\includegraphics[width=0.49\textwidth]{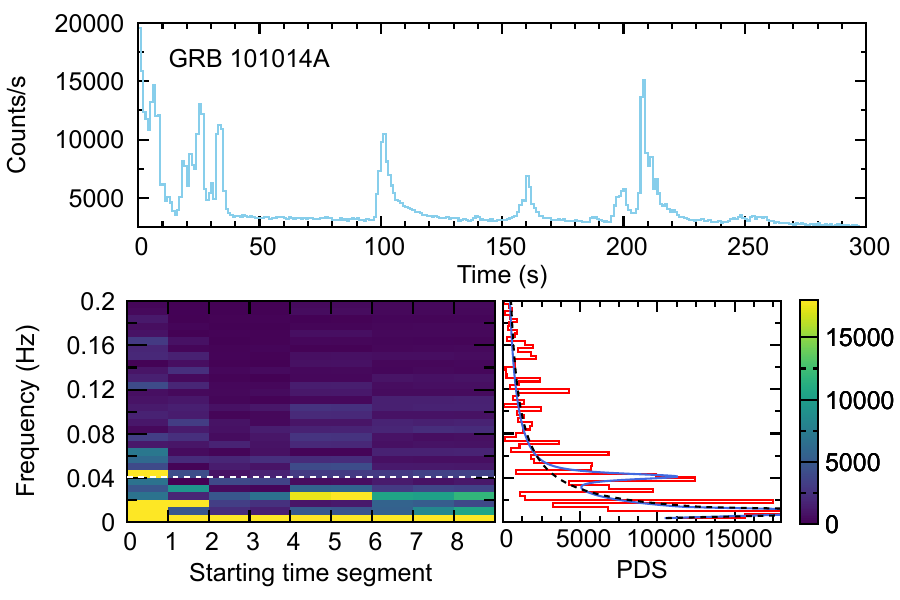}
\includegraphics[width=0.49\textwidth]{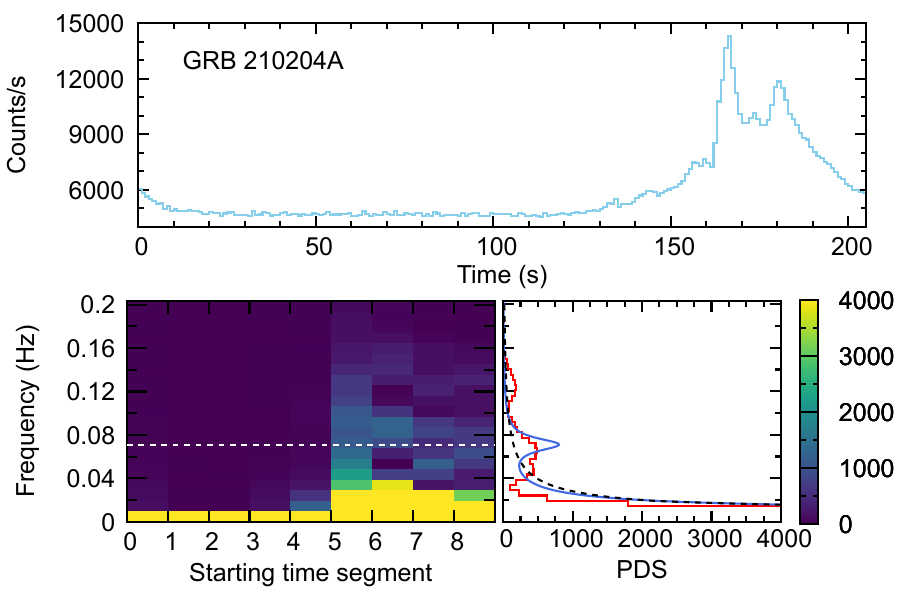}
\includegraphics[width=0.49\textwidth]{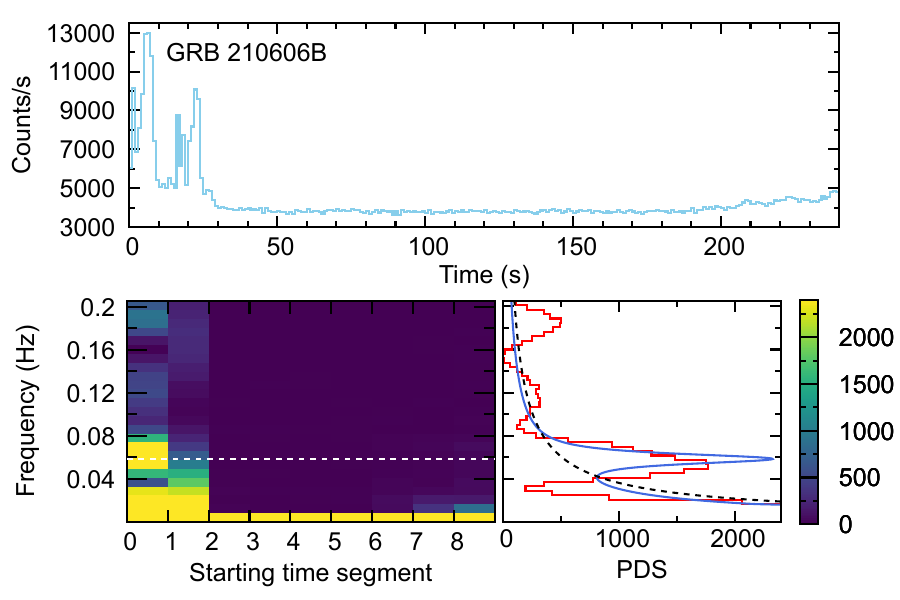}
\includegraphics[width=0.49\textwidth]{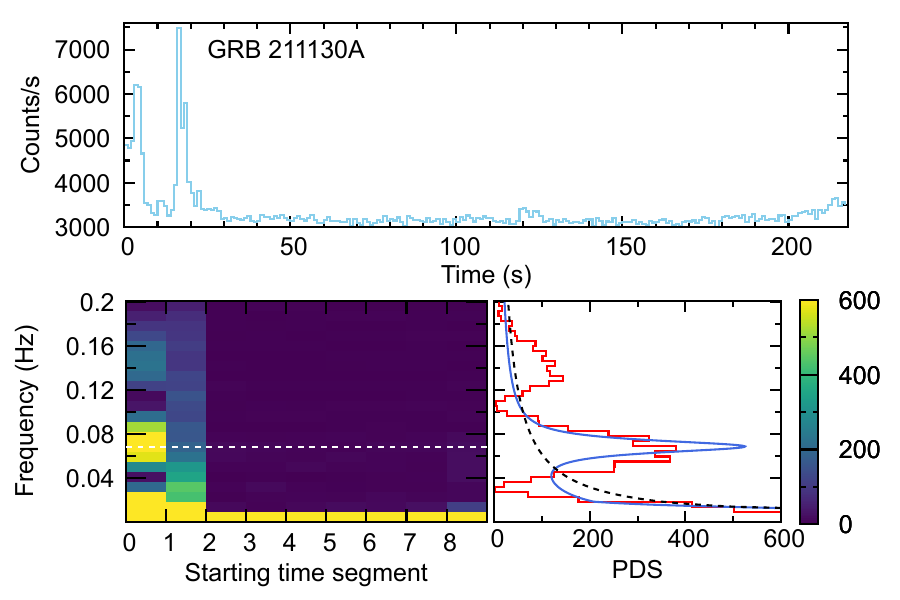}
\includegraphics[width=0.49\textwidth]{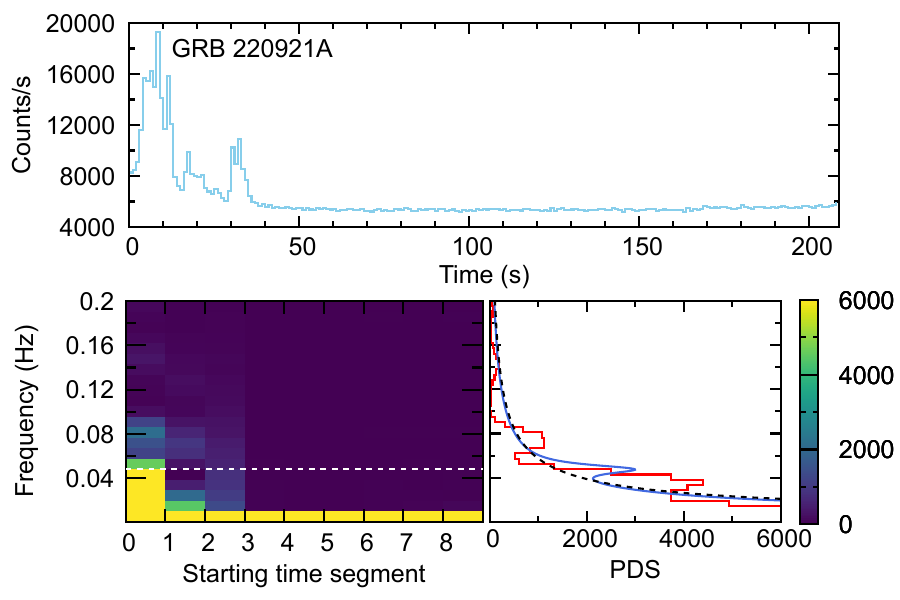}
\includegraphics[width=0.49\textwidth]{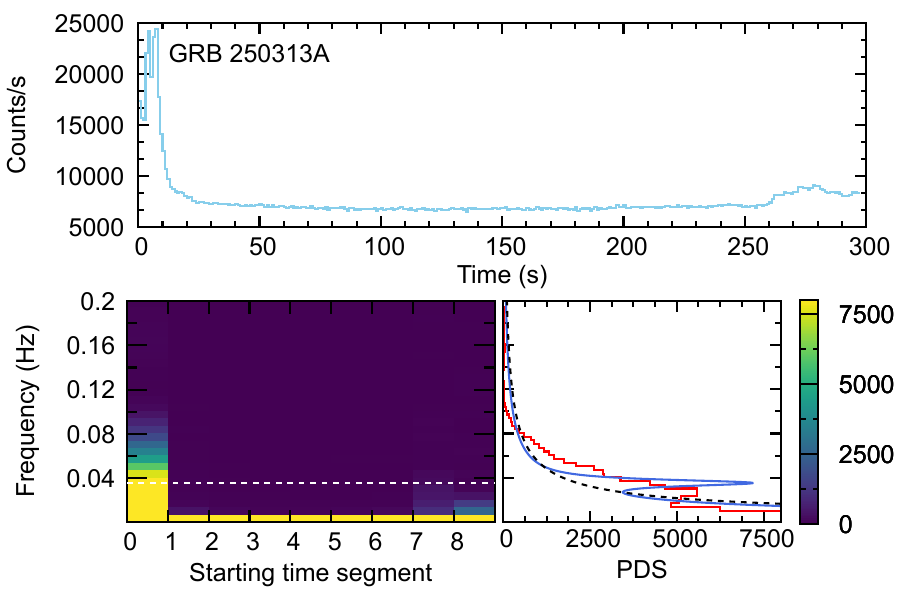}
\caption{Same as Fig.\ \ref{fig:spectrogram}, but for the 6 additional outliers shown in Fig.\ \ref{fig:rms_bayes}. 
GRBs 210204A, 210606B and 211130A show ``fake'' QPOs caused by GRB light curves showing structure consistent with two bright peaks restricted to a small portion of the light curve, resulting in an over-resolved feature in the power spectrum at a frequency $f \approx  $1/(peak separation). 
GRBs 101014A, 220921A and 250313A are the brightest QPOs in the set and show ``fake'' QPOs caused by bumps in the power spectrum with high power (due to their large number of counts) that are statistically significant, but do not correspond to a flux modulation. }
\label{fig:spectrogram_all}
\end{figure*}

In Fig.\ \ref{fig:spectrogram_all} we present the light curves, power spectra and spectrograms of the complete set of 6 outliers identified in Fig. \ref{fig:rms_bayes}, in the same format as Fig.\ \ref{fig:spectrogram}. None of these GRBs shows obvious flux modulation in the light curve, despite having extremely high Bayes factors in favor of a QPO (see Fig.\ \ref{fig:rms_bayes}). Their QPO-like features in the power spectrum appear to be caused by a double-peaked light curve (GRBs 210104A, 210606B and 211130A, see also Fig.\ \ref{fig:fake1}) or by noise fluctuations with very high power in extremely bright GRBs (GRBs 101014A, 220921A and 250313A).

This highlights one of the difficulties of QPO searches in GRBs. The analysis in the frequency domain allows us to use the Bayesian comparison method described in Sec. \ref{sec:data_bayes} to quantify the statistical significance of a QPO candidate. However, if we restrict the analysis solely to the frequency domain, there is no smoking gun that allows us to distinguish a real QPO (i.e. associated with a persistent flux modulation, as in GRB 241030A) from a ``fake'' one (whose light curve shows no obvious modulation, as seen in Fig.\ \ref{fig:spectrogram_all}). 

\begin{figure}[tbh]
\centering
\includegraphics[width=0.49\textwidth]{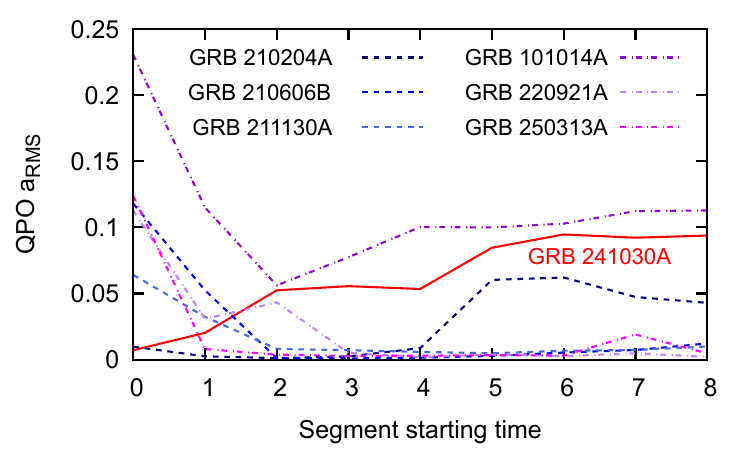}
\caption{The fractional RMS amplitude of the QPO candidate for GRB 241030A and the six GRBs shown in Figure \ref{fig:spectrogram_all}, calculated in each data segment shown in their respective spectrograms (see Figs.\ \ref{fig:spectrogram} and \ref{fig:spectrogram_all}). GRB 241030A shows increasing $a_{\rm RMS}$ that levels out at $\sim 0.1$, indicating a persistent flux modulation. Most of the GRBs with ``fake'' QPOs show $a_{\rm RMS}$ that starts high and falls fast, indicating GRBs with most of the counts at the beginning of the burst and with no persistent flux modulation. 
(For GRB 210204A, this behavior is inverted, because most of the counts appear at the end of the burst.) 
GRB 101014A is the brightest GRB among the outliers and is the only burst with $a_{\rm RMS}$ higher than GRB 241030A.}
\label{fig:a_rms_time}
\end{figure}

We present a preliminary diagnostic to address this problem in Fig.\ \ref{fig:a_rms_time}. For most of the GRBs, the time-resolved fractional RMS amplitude of the QPO peaks early in the light curve. A persistent flux modulation should correspond to a flatter $a_{\rm RMS}$ curve, as in the case of GRB 241030A (see the caption for more details). However, scaling this diagnostic for a large set of GRBs would require the determination of thresholds for interesting cases, which could introduce more ambiguity.

\section{Discussion}
\label{sec:discussion}

In this section we assume that the QPO is a genuine characteristic of the system, discuss candidate physical mechanisms, and assess their viability.
To be viable, a proposed mechanism must last for hundreds of seconds while maintaining an approximately constant period and phase coherence.

{\bf Rotation of a magnetar.}  Neutron stars have a wide range of rotation periods, from milliseconds to hours \citep{2015PhR...548....1M}.  The inferred period of $\sim$25 seconds is therefore plausible.  However, if we assume that GRB~241030A signals a core collapse and the birth of a neutron star, then there are at least two challenges: (a)~if the star rotates this slowly, can it produce a jet with a Lorentz factor $\gamma$ that is typical of gamma-ray bursts ($\gamma\sim$ hundreds), and (b)~if the star was born spinning much more rapidly, is it possible that it did so before the main emission but then maintained a nearly constant period thereafter?

{\bf Precession of a magnetar.} A related possibility is magnetar precession.  In this scenario, the observed QPO period is not the magnetar spin period itself, but a geometric modulation period associated with precession of a rapidly spinning oblate magnetar. As the magnetar precesses, its emission beam can move into and out of our line of sight, producing quasi-periodic variability. 
A possible $\sim$50 s QPO ($6\sigma$) in the early X-ray afterglow of GRB 220711B was reported by \citet{2025MNRAS.541.3787S}, and a possible $\sim$11 QPO ($3\sigma$) in the prompt gamma-ray emission of GRB 210514A was reported by \citet{2024ApJ...973..126Z}; both of which were interpreted as the precession periods of newly formed proto-magnetars.  In this scenario, the magnetar spin period is much shorter, with inferred initial periods of order a few ms, while the observed $\sim 50$ s modulation is produced by precession. However, the large magnetic fields inferred by \citet{2025MNRAS.541.3787S}, $B \sim 10^{16}$ G, imply rapid spin-down timescales, ranging from seconds to hundreds of seconds.  A rapidly spinning-down magnetar should have a lengthening precession period, so maintaining a roughly stable precession period over hundreds of seconds is challenging.

{\bf Compact object magnetospheric effects.}  Yet another possibility is QPOs originating from magnetospheric effects (e.g., the magnetospheres of neutron stars, magnetars, or accreting black holes). 
Such models include: magnetospheric plasma oscillations \citep{1995MNRAS.275..255T}, 
matter ring oscillations \citep{1999PhR...311..259W}, 
magnetospheric Alfven wave oscillations \citep{2021Natur.600..621C}, 
and magnetospheric beat-frequency mechanisms in accreting neutron stars \citep{1985Natur.316..239A,1985Natur.317..681L}.
However, as stated by \citet{2025MNRAS.541.3787S} in their discussion of the possible QPO in GRB 220711B, these scenarios are unlikely to produce QPOs in GRBs because the magnetospheric oscillation and matter ring oscillation models require a stable magnetosphere and/or matter ring to generate the oscillations.
The extremely bright and variable radiation of GRB prompt emission and the matter in the messy collapsar environment would disrupt the magnetosphere or matter rings, thus precluding the existence of a stable magnetosphere. 
Additionally, magnetospheric Alfven wave oscillations and rotation modulation models predict that observed QPOs would have frequencies from tens to hundreds of Hz, which is incompatible with these low-frequency QPOs. 

Magnetically regulated accretion has also been invoked to explain variability in prompt GRB emission. 
In the magnetically arrested disc (MAD) model of \citet{2016MNRAS.461.1045L}, magnetic flux accumulated near the black hole intermittently arrests the inner accretion flow, causing the system to transition into and out of a state in which accretion and jet power are suppressed. 
The characteristic variability timescale in this model is set by the free fall time across the arrested region, and was proposed to explain the $\sim 1$ s variability timescales seen in prompt GRB light curves. 
This is substantially shorter than the $\sim 25$ s periodicity seen in GRB 241030A. 
Additionally, even if the arrested region were large enough to produce a timescale of tens of seconds, the MAD mechanism is expected to generate episodic or stochastic pulse-to-pulse variability rather than a narrow phase-coherent QPO persisting over hundreds of seconds. 
Therefore, we consider a magnetically-arrested disc as an unlikely explanation for the QPO in GRB 241030A.

{\bf Wobbling jet.}  
Wobbling jets are often invoked to explain GRB variability. 
In 3D general relativistic magnetohydrodynamic collapsar simulations, \citet{2022ApJ...933L...9G} found that the disk-jet system can spontaneously develop a misalignment relative to the black-hole spin axis, causing the jet to wobble by tilt angle $\theta_t \sim 12^\circ$. 
Such a wobble can naturally produce intervals of gamma-ray emission.
However, this mechanism is more naturally associated with intermittency (e.g., the period of gamma-ray quiescence seen from about 50 to 100 s post-trigger) than with a narrow, phase-coherent QPO. 
Explaining the apparent $\sim$ 25 s period over hundreds of seconds requires a coherent clock, which is difficult to explain in the jet wobbling scenario, as it lacks a stable/coherent clocking mechanism.

{\bf Lense-Thirring precession of an accretion disk feeding a newly-created black hole.}  If, for example, the initial fallback of matter that created the black hole or neutron star was tilted relative to later fallback, we could imagine that the disk produced by the later fallback would have its orbital axis precess due to frame-dragging.  Numerical studies suggest that with accretion rates of Eddington or beyond, the disk might precess as a solid body \citep{2005ApJ...623..347F,2007ApJ...668..417F} rather than undergoing alignment into the black hole rotational plane.  If so, then depending on the mass and accretion rate, the precession frequency might be hundredths of a Hz.  
A potential challenge here is again to maintain a roughly constant frequency despite the expected substantial changes in the accretion rate.
Another issue is that after the period of gamma-ray quiescence (from about 50 to 100~s post-trigger), when the gamma-ray emission starts up, it appears phase-connected.  
If the accretion stopped (or was significantly reduced), the burst should not maintain its phase coherence throughout its duration, because the precession rate should decrease with decreasing accretion rate, which would thus break phase coherence which we see in the light curves (see Figure~\ref{fig:tdrss_BAT_LC}).

{\bf Compact object oscillations}.  The low observed frequency rules out a variety of compact object oscillation mechanisms. 
The expected ringdown frequency for a stellar-mass black hole is on the  order of hundreds to thousands of Hz. 
A frequency of 0.04 Hz would require a black hole mass on the order of $10^5-10^6 \ M_\odot$ (as black hole quasi-normal modes scale inversely with the BH mass), which is not compatible with a newly formed black hole from a collapsar. 
Similarly, neutron star oscillations are expected at much higher frequencies: typical f-modes occur at $\sim1.5-3$ kHz, with dampening times $<1$ s \citep{1999LRR.....2....2K,2009CQGra..26p3001B}. 
Magnetar seismic QPOs are generally observed or expected at $\gtrsim 10-30$ Hz \citep{2007AdSpR..40.1446W}.
Thus, the 0.04 Hz modulation in GRB 241030A is unlikely to be a BH or NS oscillation.

{\bf Jet collision with regularly-spaced layers of stellar ejecta in the circumburst medium}.  
If the pre-collapse star had a binary companion in an eccentric orbit, which, near periastron, became close enough to modulate the mass loss rate via interaction of stellar winds, then the circumburst medium would consist of stratified layers of stellar ejecta.
A GRB jet colliding with that stratified outflow, with an observer-frame interval compressed by relativistic effects, could lead to modulation in the emission produced by external shocks.  
For example, if the Lorentz factor $\gamma\sim 100$, then the time compression factor is $\approx 1/(2\gamma^2)$.  
For a wind with a radial outflow speed $v$ (for example, $v\sim 0.01c$ for massive main-sequence stars), an orbital period of a year in the source frame would appear as a period of $\sim (v/c)(1/2\gamma^2)\sim$ tens of seconds in the observer frame.  
Here, questions would include why, if this happens, we see it so seldom, and whether the modulation is sufficient.

\begin{figure}[tbh]
\centering
\includegraphics[width=0.49\textwidth]{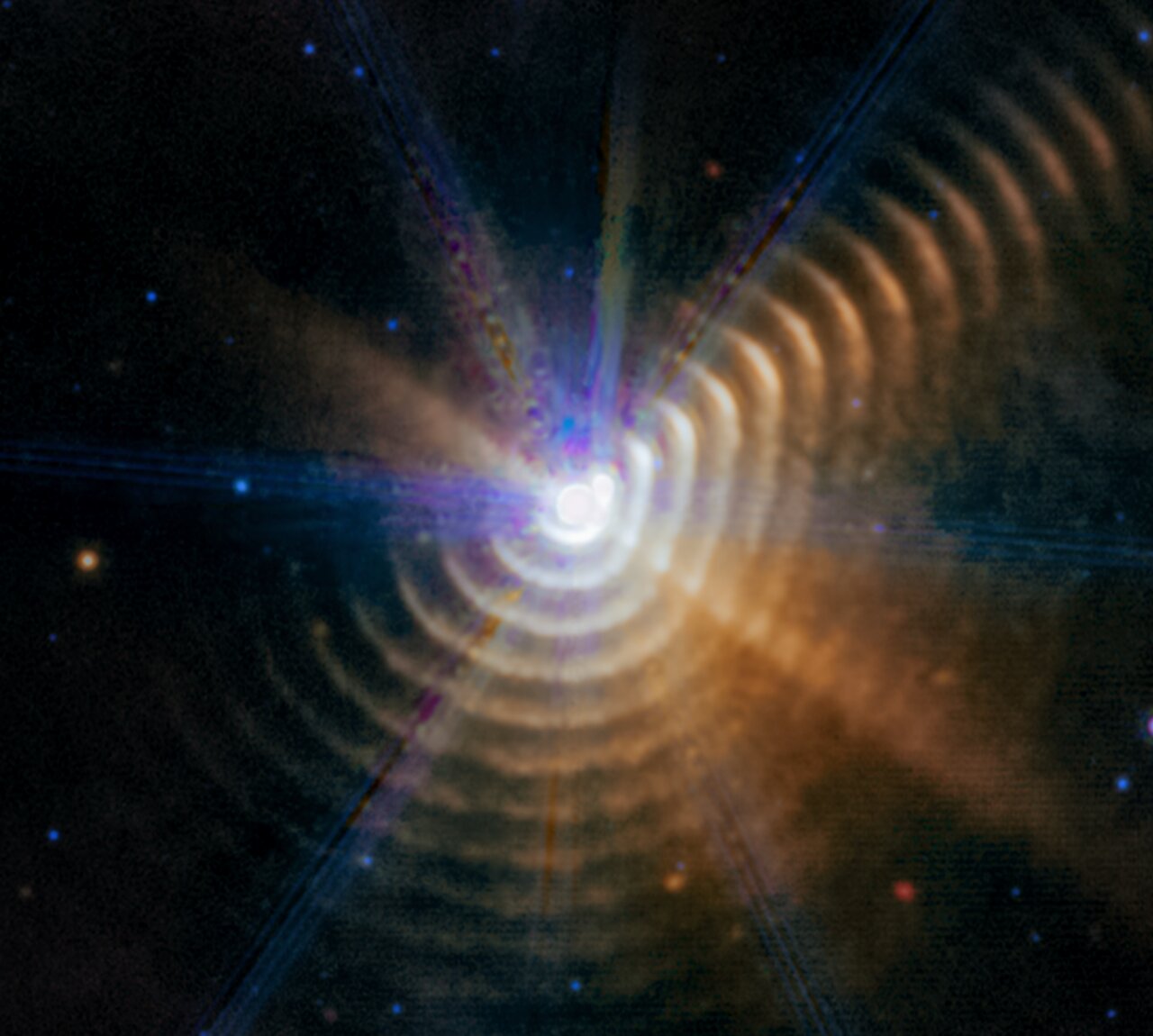}
\caption{JWST/MIRI image (false color) of WR 140 showing its concentric nested shells of dust/ejecta: a system which could produce QPO-like features in a light curve as a GRB jet passed through it, shown for illustrative purposes.  Image credit: NASA, ESA, CSA, STScI, E.~Lieb (University of Denver), R.~Lau (NSF NOIRLab), J.~Hoffman (University of Denver).}
\label{fig:WR140}
\end{figure}

Although this scenario may seem contrived, such shells of regularly-spaced stellar ejecta have been observed around Wolf-Rayet binaries in our own galaxy. 
For example, the eccentric WR + O binary system WR~140 \citep{2002ApJ...567L.137M} emits shells of dust every 8 years due to binary wind interactions at the system's periastron passage. 
Recent JWST-MIRI observations have detected at least 17 shells \citep{2022NatAs...6.1308L} from this system, which we show in Figure \ref{fig:WR140}. 
A GRB jet passing through regularly-spaced concentric ejecta shells like in this system could produce regularly-spaced peaks in the observed light curve. 
Thus, in this scenario, the QPO would serve as a probe of the circumburst medium and evolutionary history of the progenitor star, rather than a probe of the central engine.

WR~140 is an unusually ``clean'' example of this phenomenon, because its highly eccentric orbit produces discrete dust-formation episodes near periastron.  
However, ordered circumstellar structure is not unique to WR~140. 
Other dusty Wolf-Rayet binaries, including WR~104 and WR~98a, show spiral or pinwheel-like dust nebulae produced by colliding winds in binary systems \citep{1999Natur.398..487T,1999ApJ...525L..97M}, while WR~112 shows evidence for a larger-scale spiral dust structure associated with a dust-forming carbon-rich WR binary \citep{2020ApJ...900..190L}. 
Periodic or episodic dust formation has also been reported in WR~137 and WR~19 \citep{2001MNRAS.324..156W,2009MNRAS.395.2221W}. 
These examples show that WR binary interactions do produce ordered, radially structured circumstellar material even when the morphology is not as cleanly shell-like as in WR~140.
As we see no obvious challenges to this scenario, we consider this as our favored explanation.

\section{Conclusions}
\label{sec:conclusions}

We have presented a multi-wavelength timing analysis of GRB 241030A: a long GRB with an unusual light curve containing $\sim$10 peaks that are approximately aligned with a period of $\sim 12$--$13$ s.  
In the frequency domain, the GBM, BAT, and XRT data show excess power near $f_0 \simeq 0.04$ Hz, corresponding to a period of $\sim$25 s, as well as excess power near harmonics of this frequency.  
The $\sim 0.04$ Hz feature is statistically favored over a red-noise-only model in the GBM, BAT, and XRT data, and a $\sim 0.08$~Hz feature is strongly favored in the GBM and BAT data.  The UVOT data show a weak but not statistically significant excess at the same two frequencies.  
To the best of our knowledge, this is the first reported GRB QPO  detected at a consistent frequency in both soft X-rays (Swift/XRT) and in gamma-rays (Swift/BAT and Fermi/GBM).

The interpretation of this feature is not straightforward.  
The apparent QPO lies at low frequency, where GRB power spectra are strongly affected by red noise, and where the complex structure of the prompt light curve can produce QPO-like excesses in the frequency domain.  
In our comparison with a sample of long GRBs with similar durations, GRB 241030A stands out from most bursts with comparable count numbers, but the comparison also shows that large Bayes factors can be produced by light curves without persistent flux modulation.  
We therefore cannot state with certainty whether the feature in GRB 241030A is a true QPO rather than an unusually structured red-noise-dominated variability.

Nevertheless, the visual distinctiveness of the light curve, the presence of excess power in multiple instruments and at multiple energies, and the approximate consistency of the candidate frequency and its harmonics suggest that the modulation may reflect a characteristic frequency of the system.  
If so, the low frequency disfavors direct compact object oscillation modes and instead points toward a mechanism capable of modulating the jet, or its interaction with the circumburst environment, over hundreds of seconds.  
After considering all plausible explanations, we consider the most likely mechanism to be a GRB jet moving through regularly-spaced circumburst ejecta shells, produced by orbital modulation with a binary companion. 
GRB 241030A therefore provides both a promising candidate for a GRB QPO and a useful cautionary example of the difficulty of distinguishing true quasi-periodicity from complex low-frequency GRB variability.

\begin{acknowledgments}

N.K.\ and C.C.\ were supported by NASA under award number 80GSFC24M0006.
C.C.\ and M.C.M.\ were supported in part by NASA grants 80NSSC25K7108, 80NSSC25K0511, and 80NSSC26K0667.  Part of this work was performed at the Aspen Center for Physics, which is supported by National Science Foundation grant PHY-2210452. 
S.D.\ was also supported by NASA under award number 80NSSC26K0666.
N.K.\ would like to thank Sam Shilling for helpful discussions and contributions to the preliminary timing analysis of GRB 241030A.

\end{acknowledgments}

\facilities{Fermi(GBM), Swift(BAT, XRT, UVOT)}

\software{HEASoft}

\appendix

\section{BAT Data and Nearby Variable Sources}
\label{sec:appendix1}

Three relatively bright and variable X-ray sources, Cyg X-1, X-2, and X-3 were in the BAT field of view during the time of the burst. 
Because the analysis uses non mask-weighted light curves, it is possible that the light curve contains contributions from the Cygnus sources. 
To examine the potential contamination, we examined the power spectrum using the mask-weighted light curve from Cyg X-1. For a coded-mask instrument, the mask-weighted light curve only includes photons from the source location, while the non-mask-weighted light curves include photons from the entire BAT field of view. While the QPO analysis requires that we use the non-mask-weighted light curves, we examine the contribution of the power spectrum from Cyg X-1 using the mask-weighted light curves. 

\begin{figure}[tbh]
\centering
\includegraphics[width=0.45\textwidth]{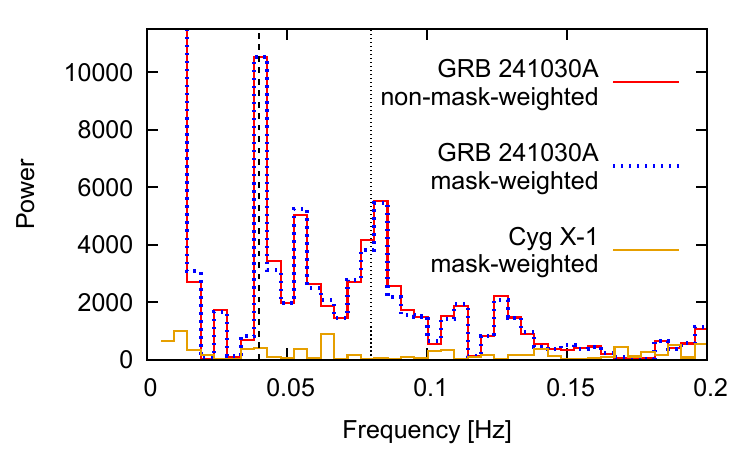}
\caption{Comparison of power spectra obtained for the non-mask-weighted and mask-weighted light curves for GRB 241030A. 
We also show the mask-weighted light curve of Cyg X-1 (a bright, variable source in the BAT field of view) in the duration of the GRB. 
The plot shows that Cyg X-1 is not noticeably contaminating the non-mask-weighted power spectra of GRB 241030A.
For our comparison we use the power spectral normalization of \citet{1975ApJS...29..285G} for all three curves, even though mask-weighting means that Poisson statistics no longer applies.}
\label{fig:compare_X1}
\end{figure}

Figure~\ref{fig:compare_X1} shows the comparison of power spectra from the mask-weighted light curves of Cyg X-1 to both the non-mask-weighted and mask-weighted light curve of GRB 241030A. As seen in the figure, the power spectra from the non-mask-weighted and mask-weighted light curves show very similar behavior, and Cyg X-1 contributes very little to the overall power spectrum.

Since we do not find any significant contribution in the power spectrum from Cyg X-1, we expect even smaller contributions from Cyg X-2 and X-3, as these sources were much fainter than Cyg X-1 when they were in the BAT field of view ($7 \sigma$ for Cyg X-1 versus lower than $3 \sigma$ for Cyg X-2 and X-3 in the post-slew image during the majority of the GRB emission.)

\bibliography{bibliography}{}
\bibliographystyle{aasjournalv7}

\end{document}